\PassOptionsToPackage{table, dvipsnames}{xcolor}
\documentclass[letterpaper,twocolumn,10pt]{article}
\usepackage{usenix2020_09}

\usepackage{tikz}
\usepackage{amsmath, amsfonts}
\usepackage{graphicx}
\usepackage{textcomp} 
\usepackage[T1]{fontenc}
\usepackage[utf8]{inputenc}
\usepackage{xspace}
\usepackage{multirow}
\usepackage{diagbox}
\usepackage{booktabs}
\usepackage{caption}
\usepackage{subcaption}
\usepackage{float}
\usepackage{makecell}
\usepackage{paralist}

\usepackage{enumitem}
\usepackage{fvextra}
\usepackage[skins]{tcolorbox}
\usepackage[ruled,lined,linesnumbered]{algorithm2e}
\usepackage{algorithmicx}
\usepackage[numbers,sort&compress]{natbib}
\usepackage{pifont}
\usepackage[pagebackref=true,breaklinks=true,colorlinks,bookmarks=false]{}
\usepackage{hyperref}
\hypersetup{
  colorlinks,
  linkcolor={RoyalBlue},
  citecolor={PineGreen},
  urlcolor={Bittersweet}
}
\definecolor{mycolor}{RGB}{246, 196, 188}
\definecolor{codegray}{rgb}{0.4, 0.4, 0.4}
\usepackage{soul}
\sethlcolor{mycolor}
\newcommand{\hlcolor}[2]{
  \sethlcolor{#1}
  \hl{#2}
  \sethlcolor{mycolor}
}
\colorlet{lightgray}{gray!20}

\setlist[itemize]{leftmargin=*}

\SetCommentSty{mycommfont}

\def\X#1{\ding{\numexpr181+#1}}

\definecolor{revision}{RGB}{0,0,255}

\newcommand{\revstart}{\begin{color}{revision}}
\newcommand{\revend}{~\!\!\end{color}}

\newcommand{\mypara}[1]{\smallskip \noindent\textbf{#1.} \xspace}

\newcommand{\method}{\ensuremath{\mathsf{GradEscape}}\xspace}
\newcommand{\aigt}{{AIGT}\xspace}
\newcommand{\sptoken}[1]{{\fontfamily{qcr}\selectfont#1}}
\newcommand{\block}[1]{%
  \raisebox{\dimexpr(\fontcharht\font`X-1em)/2}{\rule{0.5em}{#1\dimexpr1em/8}}%
}
\DeclareUnicodeCharacter{2581}{\block{1}}
\DeclareUnicodeCharacter{2582}{\block{2}}
\DeclareUnicodeCharacter{2583}{\block{3}}
\DeclareUnicodeCharacter{2584}{\block{4}}
\DeclareUnicodeCharacter{2585}{\block{5}}
\DeclareUnicodeCharacter{2586}{\block{6}}
\DeclareUnicodeCharacter{2587}{\block{7}}
\DeclareUnicodeCharacter{2588}{\block{8}}

\newtcolorbox{mybox}[2][]{text width=0.95\columnwidth, fontupper=\normalsize,
fonttitle=\bfseries\sffamily\normalsize, colbacktitle=codegray, enhanced,
boxed title style={sharp corners}, top=4pt, bottom=2pt, left=2pt, right=2pt,
title=#2, colback=gray!10}

\usepackage{etoolbox}
\makeatletter
\patchcmd{\hyper@makecurrent}{%
    \ifx\Hy@param\Hy@chapterstring
        \let\Hy@param\Hy@chapapp
    \fi
}{%
    \iftoggle{inappendix}{
        \@checkappendixparam{chapter}%
        \@checkappendixparam{section}%
        \@checkappendixparam{subsection}%
        \@checkappendixparam{subsubsection}%
        \@checkappendixparam{paragraph}%
        \@checkappendixparam{subparagraph}%
    }{}%
}{}{\errmessage{failed to patch}}

\newcommand*{\@checkappendixparam}[1]{%
    \def\@checkappendixparamtmp{#1}%
    \ifx\Hy@param\@checkappendixparamtmp
        \let\Hy@param\Hy@appendixstring
    \fi
}
\makeatletter

\newtoggle{inappendix}
\togglefalse{inappendix}

\apptocmd{\appendix}{\toggletrue{inappendix}}{}{\errmessage{failed to patch}}

\usepackage{xr}
\makeatletter

\newcommand*{\addFileDependency}[1]{
\typeout{(#1)}
\@addtofilelist{#1}
\IfFileExists{#1}{}{\typeout{No file #1.}}
}\makeatother

\usepackage[available,functional,reproduced]{usenixbadges}

\begin{document}

\title{\Large \bf  \method: A Gradient-Based Evader Against AI-Generated Text Detectors}

\author{
{\rm Wenlong Meng$^{\dagger}$ \quad Shuguo Fan$^{\dagger}$ \quad Chengkun Wei$^{\dagger, \ast}$ \quad Min Chen$^{\wr}$ \quad Yuwei Li$^{\natural \S}$}\\ {\rm Yuanchao Zhang$^{\ddagger}$ \quad Zhikun Zhang$^{\dagger, \ast}$ \quad Wenzhi Chen$^{\dagger}$}\\
{\small$^{\dagger}$Zhejiang University \quad $^{\wr}$Vrije Universiteit Amsterdam \quad $^{\natural}$National University of Defense Technology} \\ {\small $^{^\S}$Anhui Province Key Laboratory of Cyberspace Security Situation Awareness and Evaluation \quad 
$^{\ddagger}$Mybank, Ant Group}\\
{\tt\small \{mengwl, fanshuguo, weichengkun, zhikun, chenwz\}@zju.edu.cn}\\
{\tt\small m.chen2@vu.nl, liyuwei@nudt.edu.cn, yuanchao.zhang@mybank.cn}
}

\maketitle
\let\oldthefootnote\thefootnote
\renewcommand{\thefootnote}{}
\footnotetext{$^{\ast}$Corresponding authors.}
\renewcommand{\thefootnote}{\oldthefootnote}

\begin{abstract}

In this paper, we introduce \method, the first gradient-based evader designed to attack AI-generated text (AIGT) detectors. \method overcomes the undifferentiable computation problem, caused by the discrete nature of text, by introducing a novel approach to construct weighted embeddings for the detector input. It then updates the evader model parameters using feedback from victim detectors, achieving high attack success with minimal text modification. To address the issue of tokenizer mismatch between the evader and the detector, we introduce a warm-started evader method, enabling \method to adapt to detectors across any language model architecture. Moreover, we employ novel tokenizer inference and model extraction techniques, facilitating effective evasion even in query-only access.

We evaluate \method on four datasets and three widely-used language models, benchmarking it against four state-of-the-art AIGT evaders. Experimental results demonstrate that \method outperforms existing evaders in various scenarios, including with an 11B paraphrase model, while utilizing only 139M parameters. We have successfully applied \method to two real-world commercial AIGT detectors. Our analysis reveals that the primary vulnerability stems from disparity in text expression styles within the training data. We also propose a potential defense strategy to mitigate the threat of AIGT evaders. We open-source our \method for developing more robust \aigt detectors.\footnote{Code is available at \url{https://doi.org/10.5281/zenodo.15586856}.}

\end{abstract}

\section{Introduction}
\label{sec:intro}

The past years have witnessed the powerful capability of large language models (LLMs) in natural language generation (NLG) tasks~\cite{openai2023gpt4, anil2023palm, chatgpt, copilot}.
However, AI-generated text (\aigt) may be used for malicious purposes, such as fabricating plausible yet false information~\cite{leite2023detecting}, automating the production of spam and phishing email~\cite{abdelnabi2023not}, and creating biased or discriminatory content~\cite{yeh2023evaluating}.
Furthermore, many studies have revealed that AIGT is difficult to distinguish by humans ~\cite{sadasivan2023can, chen2023can}.

To mitigate the misuse of \aigt, researchers have proposed a number of methods for \aigt detection~\cite{zhong2020neural, tian2023multiscale, kumari2023demasq, hu2023radar, krishna2024paraphrasing, kirchenbauer2023watermark}.
The most widely used approach is fine-tuning a language model to build a binary classifier, which has been demonstrated effective by many studies~\cite{tian2023multiscale, guo2023close, liu2023check}.
Owing to its ease of employment and independence from the generation process, LM fine-tuning-based detector has become the preferred approach for real-world detection products, including Sapling~\cite{sapling}, Scribbr~\cite{scribbr}, and GPTZero~\cite{gptzero}.

In parallel, there has been growing interest in developing evasion techniques designed to generate adversarial texts that bypass detection. 
Evasion techniques are essential for uncovering vulnerabilities in AI detection models, improving their robustness, and simulating real-world attacks~\cite{fawzi2018adversarial, zhang2024does}. These techniques enable researchers to stress test detection systems, enhance resilience against adversarial inputs, and anticipate future threats to build more secure AI systems.
Existing evasion techniques fall into two categories: perturbation-based and paraphrase-based methods.
Perturbation-based evaders modify keywords in the text that significantly influence detection outcomes~\cite{pu2023deepfake, jin2020bert}, while paraphrase-based evaders employ LMs to rewrite the text~\cite{krishna2024paraphrasing, sadasivan2023can}.

However, our analysis reveals notable shortcomings in current \aigt evaders, particularly regarding compromised text integrity and quality (see \autoref{sec:eval_open_model_attack}).
Specifically, perturbation-based methods~\cite{pu2023deepfake, jin2020bert} often degrade the natural flow of the text, leading to awkward or grammatically incorrect sentences.
Paraphrase-based strategies~\cite{krishna2024paraphrasing, sadasivan2023can}, while maintaining grammatical correctness, tend to make extensive and uncontrollable changes to the original text's meaning or style, resulting in a loss of the original intent and key details.

Gradient-based methods offer a promising solution to the issues of text integrity and quality.
For example, in the image domain, gradient-based adversarial attacks are commonly employed to create imperceptible perturbations that alter a model's output~\cite{xiao2018generating, zhang2022adversarial}.
\textit{However, the discrete nature of text presents a significant challenge for applying such methods in the text domain because of undifferentiable computation.}

\mypara{Our Contributions} 
To fill this gap, in this paper, we propose the first gradient-based evader, named \method.
\method fine-tunes a \textit{sequence-to-sequence} (seq2seq) model and transforms it into a paraphrase, which enables \aigt to be paraphrased in such a way that can be recognized by detectors as human-generated text while maintaining syntactic and semantic integrity.

To solve the undifferentiable problem, we intercept the token probability output of the seq2seq model and use the token probabilities at each position to weight the embedding dictionary of the victim detector.
These weighted embeddings can then be fed into the detector.
To address the issue of tokenizer mismatch between the evader and the detector, we propose a \textit{warm-started evader} method, where a pretrained seq2seq model is built using the same tokenizer as the detector model.
These methods overcome the challenge posed by the discrete nature of text by operating in a continuous embedding space, enabling seamless integration of gradient-based techniques.

We formulate our \method as an optimization problem with two objectives: (1) the \textit{function constraint}, which focuses on fooling the victim detector, and (2) the \textit{consistency constraint}, aimed at maintaining the text integrity and quality.
We implement these two constraints with dedicated losses and transform the evader model training to an unsupervised manner that requires only \aigt rather than human-generated text as training data.
Specifically, to enforce function constraint, we feed the probability output of the seq2seq model into the frozen victim detector, subsequently optimizing the seq2seq model to elicit human-label predictions from the detector.
For the consistency constraint, we employ pairwise cross-entropy loss to maintain syntactic similarity and leverage text encoder models to ensure semantic coherence.

In the opaque model scenario, a more strict and practical situation where an attacker has only query access to the victim detector, we develop an \textit{opaque model attack}.
This attack involves constructing a \textit{surrogate detector} and then training the evader against it.
We determine the detector’s architecture through a novel \textit{tokenizer inference attack}, which speculates a model's tokenizer to infer its architecture.
Then we construct similar parameters via \textit{model extraction attacks}~\cite{tramer2016stealing, orekondy2019knockoff, chen2023d}.
To our knowledge, \textit{we are the first to exploit inherent vulnerabilities of LM tokenizers to inference model privacy.}

We conduct experiments on 4 \aigt detection datasets and compare \method with 4 state-of-the-art evaders.
When the ROUGE score, a widely used text consistency metric, is set to 0.9, our method consistently achieves higher evasion rates across all datasets, surpassing the performance of other evaders, including one powered by an 11B LLM.
\method achieves this with only 139M parameters.
In the opaque model attack scenario, using just 2,000 queries, our \method achieves evasion rates that are 90\% as effective as when the attacker has full knowledge of the victim detector.
We also apply \method against two real-world commercial \aigt detectors.
With a query cost of \$10, \method achieves an average evasion rate of 0.617.

Furthermore, to mitigate the threat posed by \method, we propose a novel black-box, detector-independent defense that removes the subtle modifications made by evaders.
Our defense leverages an off-the-shelf LLM to paraphrase input text before detection, ensuring that all inputs share a consistent expression style.
With this defense, we successfully reduce the evasion rate to 20\%.
In essence, our contributions can be summarized as follows:
\begin{compactitem}[$\bullet$]
    \item We present \method, the first gradient-based evader, for attacking \aigt detectors.
    \method leverages victim detectors' gradient to update its model parameters, which allows \method to achieve stronger attacks with a small model size.
    At the same time, we introduce a warm-started evader technique, making \method adaptable for attacking detectors built on arbitrary LM architecture.
    \item We introduce the opaque model attack, enabling the adaptation of \method to scenarios where the attacker has only query access to the victim detector.
    Opaque model attack comprises a novel tokenizer inference technique that allows for probing the detector's model architecture through crafting dedicated queries.
    \item We evaluate \method on 4 datasets against 3 popular LMs.
    We reproduce 4 state-of-the-art \aigt evaders and compare them with \method. Experimental results demonstrate that our \method outperforms existing evaders in various scenarios.
    Additionally, we successfully applied our \method against two real-world commercial \aigt detectors.
    \item We find that the threat arises from differences in text expression style within the training set. 
    To address this, we propose a novel, back-box, detector-independent defense that neutralizes subtle modifications made by evaders.
\end{compactitem}

\section{Preliminaries}
\label{sec:preliminaries}

\subsection{Language Models}

A language model is a probabilistic model of natural languages.
It is usually a Transformer~\cite{vaswani2017attention} model trained on large amounts of raw text in a self-supervised fashion.
This process is generally referred to as \textit{pretraining} to distinguish it from downstream training for adapting specific tasks.
Based on the approach of pretraining, LMs can be categorized into three types: causal language models, masked language models, and sequence-to-sequence models, as shown in~\autoref{tab:language_models}.

\begin{table}[t]
\centering
\caption{Language model taxonomy.}
\resizebox{\columnwidth}{!}{
\begin{tabular}{lccc}
\toprule
\textbf{LM Type}
& \textbf{Architecture}
& \textbf{Applied Tasks}
& \textbf{Roles} \\ 
\midrule
CLM
& Decoder Only
& NLG, NLU
& Generator, Detector \\ 

MLM
& Encoder Only 
& NLU
& Detector \\

Seq2seq
& Encoder-Decoder 
& NLG 
& Evader \\
\bottomrule
\end{tabular}
}
\label{tab:language_models}
\end{table}

\mypara{Casual Language Models (CLMs)}
During pretraining, CLMs predict the next token based on the preceding input.
The loss function for a given token sequence $\left\{x_1, x_2, \ldots, x_n\right\}$ and model parameters $\theta$ is:
\begin{equation}
    L_\theta=-\sum_{i=1}^{n-1} \log \operatorname{P_\theta}\left(x_{i+1} \mid x_1, x_2, \ldots, x_i\right)
\end{equation}
CLMs consist only of the decoder part of the Transformer, featuring a unidirectional attention mechanism.
CLMs possess strong text generation capabilities~\cite{radford2018improving, radford2019language}.

\mypara{Masked Language Models (MLMs)}
Represented by BERT~\cite{devlin2018bert}, MLMs aim to minimize the loss of masked token prediction.
A certain proportion of tokens (generally 15\%) are randomly selected and masked in the input sequence.
The model then predicts these masked tokens with a loss:
\begin{equation}
    L_\theta=-\sum_{x_m \in \mathcal{M}} \log \operatorname{P_\theta}\left(x_m \mid x_1, x_2, \ldots, x_{m-1}, x_{m+1}, \ldots, x_n\right),
\end{equation}
where $\mathcal{M}$ is the set of masked tokens.
MLMs utilize only the encoder part of the Transformer and feature a bidirectional attention mechanism.
They are primarily used for \textit{natural language understanding} (NLU) tasks.
With a similar number of parameters, MLMs demonstrate superior performance on NLU tasks compared to CLMs~\cite{devlin2018bert, liu2019roberta}.

\mypara{Sequence-to-Sequence Models (Seq2seq Models)}
Seq2seq models have two inputs: a source sequence $\left\{ x_1, x_2, \ldots, x_n \right\}$ and a target sequence $\left\{ y_1, y_2, \ldots, y_k \right\}$ during pretraining.
The training objective is to minimize the prediction loss of the target sequence:
\begin{equation}
    L_\theta=-\sum_{j=1}^k \log \operatorname{P_\theta}\left(y_j \mid y_1, y_2, \ldots, y_{j-1}, \mathbf{C}\right),
\end{equation}
where $\mathbf{C}=\operatorname{E_\theta}\left( x_1, x_2, \ldots, x_n \right)$ is the context vector generated by the encoder part of the seq2seq model.
Seq2seq models consist of both the encoder and decoder parts of the Transformer.
While CLMs are primarily used for general generation tasks, seq2seq models are mainly utilized for specific generation tasks, such as machine translation~\cite{xue2020mt5} and summarization~\cite{akiyama2021hie}.

\subsection{\aigt Detection}

AIGT detection techniques can be divided into \textit{post-hoc} and \textit{proactive} methods, depending on whether they require interaction with the generation process.
Proactive methods require actions during generation, while post-hoc methods do not.

\mypara{Post-hoc Detectors}
The most common post-hoc detector is a binary classifier obtained by fine-tuning an LM.
We refer to this type of detector as a deep learning-based detector.
This simple yet effective method has been validated by various studies~\cite{ippolito2020automatic, guo2023close, he2023mgtbench}.
Beyond vanilla fine-tuning, researchers proposed improvements.
For instance, CheckGPT~\cite{liu2023check} freezes the LM while adding a BiLSTM head to enhance training efficiency;
MPU~\cite{tian2023multiscale} employs a multi-scaling module and \textit{positive unlabeled} (PU) loss to increase detection accuracy for short texts; DEMASQ~\cite{kumari2023demasq} utilizes an energy-based model that involves a regulation of drumhead vibrations.
Another type of post-hoc detector is statistic-based, which trains regression models based on statistical information from the text.
For example, GLTR~\cite{gehrmann2019gltr} tracks the probability of each token's occurrence;
DetectGPT~\cite{mitchell2023detectgpt} perturbs the target text and then compares changes in perplexity.
Deep learning-based detectors generally have higher detection accuracy than statistic-based detectors, but require a large training set.
Statistic-based detectors perform better in few-shot scenarios.

\mypara{Proactive Detectors}
Unlike post-hoc detectors, proactive detectors require modifications to the LLM inference process.
Therefore, this type of detector typically needs maintenance from the LLM service provider, who then provides a detection API to users.
There are two main types of proactive detectors: watermark-based and retrieval-based.
Watermark-based detectors inject invisible watermarks into the text during generation and then measure the strength of these watermarks during detection~\cite{abdelnabi2021adversarial, he2022protecting, kirchenbauer2023watermark, fairoze2023publicly}.
For example, KGW~\cite{kirchenbauer2023watermark} randomly selects a set of ``green'' tokens before a word is generated and then softly promotes the probability of green tokens during sampling.
For detection, KGW calculates z-scores based on the number of green tokens.
Retrieval-based detector~\cite{krishna2024paraphrasing} stores historically generated text in a database.
Later, text samples are evaluated by matching against this database.

In this paper, we focus on post-hoc detectors since they have been employed by real-world products.
Watermark-based detectors jeopardize text quality and are extremely vulnerable to paraphrasing attacks~\cite{krishna2024paraphrasing, sadasivan2023can}.
Retrieval-based detectors require the storage of user data, which raises data privacy concerns and violates the European GDPR~\cite{voigt2017eu} law.
Furthermore, storing all historical data is impractical, considering the vast amount of text generated daily.
OpenAI once stated that GPT3 alone produces 4.5 billion words per day.\footnote{\url{https://openai.com/blog/gpt-3-apps}}

\subsection{Evasion Attacks}

\aigt detection is a combat.
The adversary may be aware of the existence of the detector and utilize an evader to bypass the detection by editing the generated text without changing its meaning.
Existing evasion techniques lie in two main lines: perturbation-based and paraphrase-based.

\mypara{Perturbation-based Evaders}
These evaders alter specific words in the target text, aiming to remove important words that trigger \aigt detection.
Random perturbation (RP)~\cite{pu2023deepfake} replaces random words in the text with synonyms that preserve the semantic content.
DFTFooler~\cite{pu2023deepfake} found that \aigt usually has a lower perplexity than human-generated text.
Therefore, DFTFooler replaces the top-$N$ most confidently predicted words according to a backend LM.
Different from the above two black-box evaders, TextFooler~\cite{jin2020bert} is a gray-box evader necessitating queries to the target detector.
TextFooler finds important words to replace and the best replacement words by querying the detector model.
Since TextFooler utilizes detector information, it performs a better attack performance than RP and DFTFooler when perturbing the same number of words.
However, in our experiment, TextFooler needs 500-700 queries to complete one sample attack, which makes it impractical in real-world settings.

\mypara{Paraphrase-based Evaders}
These evaders paraphrase the target text.
They assume that distinctive patterns in the output of LLMs are the primary factors that trigger \aigt detection.
DIPPER~\cite{krishna2024paraphrasing} is an 11B seq2seq model fine-tuned on PAR3 dataset~\cite{thai2022exploring}, which contains multiple English translations of non-English novels aligned at a paragraph level.
DIPPER can paraphrase a paragraph of text with specific settings for lexicon diversity and order diversity.
SentPara~\cite{sadasivan2023can} is a lightweight paraphrase-based evader.
The process begins by dividing the input text into individual sentences. Following this, SentPara applies a pre-existing sentence-level paraphrase model to rephrase each sentence sequentially. In the final step, it reassembles these paraphrased sentences into a coherent whole.
SentPara only requires a model with 222M parameters to execute the attack, which is 1/50 the size of DIPPER.

\mypara{Flaws of Existing Evaders}
Perturbation-based evaders' word-changing strategy can significantly reduce the fluency and readability of the text.
Paraphrase-based evaders' modification magnitude is hard to control.
Although DIPPER provides lexicon and order knobs, there are only five settings to choose from.
Some researchers also proposed attack methods involving the addition of special symbols in the text~\cite{cai2023evade, gonzoknows2023bypass}.
However, these special symbols can be easily filtered out, making them difficult to use in real-world scenarios.

\section{Threat Model}
\label{sec:threat_model}

\begin{figure}
    \centering
    \includegraphics[width=\columnwidth]{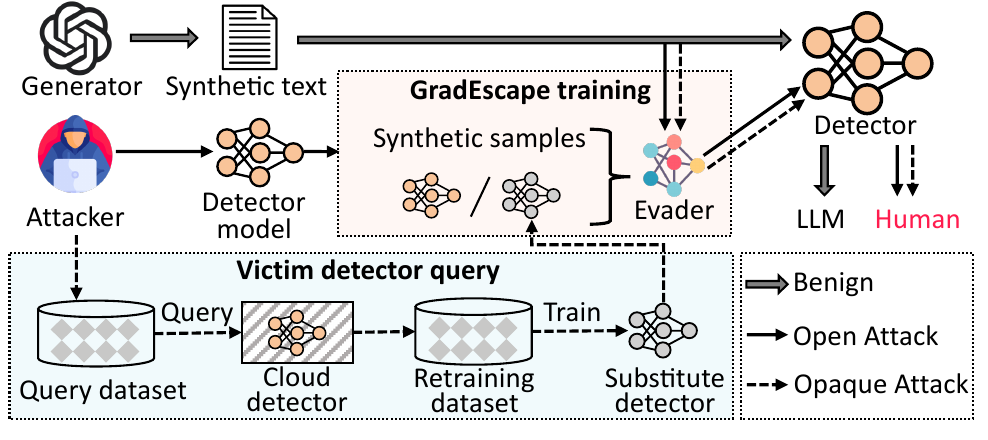}
    \caption{Attack scenarios.}
    \label{fig:threat_model}
\end{figure}

\mypara{Attacker's Goal}
We assume that the attacker is an LLM user and intends to apply the LLM-generated content in environments equipped with \aigt detectors.
The attacker aims to employ an automated evader to modify the LLM-generated text in a way that circumvents detection by the \aigt detector.
Simultaneously, the attacker wishes for the evader's modifications not to compromise the syntactic and semantic integrity and fluency of the text.
For instance, consider a scenario in which the attacker is a fraudster who is using the LLM to craft phishing emails.
Aware of the \aigt detection feature in the victim's email system, the attacker may employ an evader to alter the generated phishing email, allowing it to bypass the detection mechanism.

\mypara{Attacker's Capabilities}
The attacker can access public text datasets, such as WikiText~\cite{merity2016pointer}, and open-source LMs, such as BERT~\cite{devlin2018bert} and BART~\cite{lewis2019bart}.
We require the attacker to possess a \textit{synthetic dataset}, which contains solely LLM-generated text.
Note that, unlike detector training, which requires both LLM-generated and human-generated text, \method only needs LLM-generated samples.
Collecting the synthetic dataset is trivial for the attacker, as they can generate it using their own LLM.
For the accessibility to the victim detector, we consider two scenarios: \textit{open model attack} and \textit{opaque model attack}, as depicted in~\autoref{fig:threat_model}.

In open model attack scenarios, we assume the attacker knows the victim detector's model architecture and parameters.
This situation occurs with open-source detectors.
For instance, OpenAI released their GPT2 output detector~\cite{solaiman2019release} on Hugging Face.
Open model attacks are also practical when the attacker knows the detector's architecture and training set.

In the opaque model attack, we further illustrate that \method is also effective when the attacker can only query the victim detector, e.g., via an API, and obtain the query results.
The attacker has a query budget $N_q$ that limits the maximum number of queries that can be made.
We assume that the attacker possesses $N_q/2$ human-generated samples and $N_q/2$ AI-generated samples for querying.
By querying, the attacker can deduce the victim detector's model architecture and use query results to train a substitute model.
As a specific LM often corresponds to a particular tokenizer, the attacker only needs to identify the tokenizer used by the detector to infer its model architecture.
The attacker can craft inputs that are tokenized identically under the specific tokenizer but differently under others and then query the victim detector with these inputs.
By observing whether the detector's response scores differ, the attacker can determine whether the detector is utilizing that specific tokenizer.
We elaborate the tokenizer inference attack in~\autoref{sec:gray-box}.
Based on the responses from queries, the attacker can construct a retraining dataset to train a surrogate victim model and then use the surrogate model and the synthetic dataset to train the \method evader.

\section{\method Methodology}
\label{sec:methodology}

In this section, we will detail how to employ \method to attack open models.
We will elaborate on how \method handles opaque models in~\autoref{sec:gray-box}.

\subsection{Overview}

The goal of \aigt evasion is to revise \aigt to bypass detection while maintaining as much syntactic and semantic integrity as possible.
We can deem an evader as a transformation $\mathcal{F}_\theta$ whose input and output are both texts, where $\theta$ represents its model parameters.
Formally, given the victim detector $D_V$ and input $\left\{x_i\right\}_{i=1}^n$, the training objective of the evader model can be expressed as:
\begin{align}
\min\quad & \mathcal{L}_{D_V}\left(\mathcal{F}_\theta \left( \left\{x_i\right\}_{i=1}^n \right), y_{\text{human}} \right),\label{eq:attack_goal}  \\
\text{s.t.}\quad & \mathrm{dis}_{syn} \left( \mathcal{F}_\theta \left( \left\{x_i\right\}_{i=1}^n \right), \left\{x_i\right\}_{i=1}^n \right) < T_{syn}, \label{eq:syntactic_constraint} \\
& \mathrm{dis}_{sem} \left( \mathcal{F}_\theta \left( \left\{x_i\right\}_{i=1}^n \right), \left\{x_i\right\}_{i=1}^n \right) < T_{sem}.\label{eq:semantic_constraint}
\end{align}
Here, $\mathcal{L}_{D_V}$ is the cross-entropy loss of detector $D_V$; $\mathrm{dis}_{syn}$ and $\mathrm{dis}_{sem}$ measure the \textit{syntactic distance} and \textit{semantic distance} between the input and output of $\mathcal{F}_\theta$, respectively, while $T_{syn}$ and $T_{sem}$ represent the \textit{syntactic threshold} and \textit{semantic threshold}, respectively.
$T_{syn}$ constrains the extent to which $\mathcal{F}_\theta$ can alter the text's structure, while $T_{sem}$ ensures that $\mathcal{F}_\theta$ does not change the meaning of the text significantly.
Since seq2seq models outperform CLMs on specific NLG tasks with the same parameter size~\cite{du2022understanding, li2022learning, sun2022unified}, we choose the seq2seq model as our evader $\mathcal{F}$.

A significant challenge in evader training is the lack of parallel data, where each input text is paired with a corresponding output text, necessary for supervised fine-tuning of $\mathcal{F}_\theta$.
Constructing such a parallel dataset entails a lot of manual effort.
To solve this dilemma, we transform evader training into an unsupervised learning method by reformulating~\autoref{eq:attack_goal} to~\autoref{eq:semantic_constraint} into distinct optimization objectives.
To implement the function constraint (\autoref{eq:attack_goal}), we pass the gradient of the detector's classification loss to the evader.
To enforce the consistency constraint (\autoref{eq:syntactic_constraint} and~\autoref{eq:semantic_constraint}), we introduce two losses to increase the syntactic and semantic similarity between the paraphrased and the input text.

The above training strategy is also part of our opaque model attacks, where the attacker does not have access to the internal parameters or architecture of the victim detector.
In this scenario, once a substitute detector is obtained, the attacker can use the substitute detector as $D_V$ to train evaders in the same manner.
We elaborate on the process of obtaining substitute detectors in~\autoref{sec:gray-box}.

\begin{figure}
    \centering
    \includegraphics[width=\columnwidth]{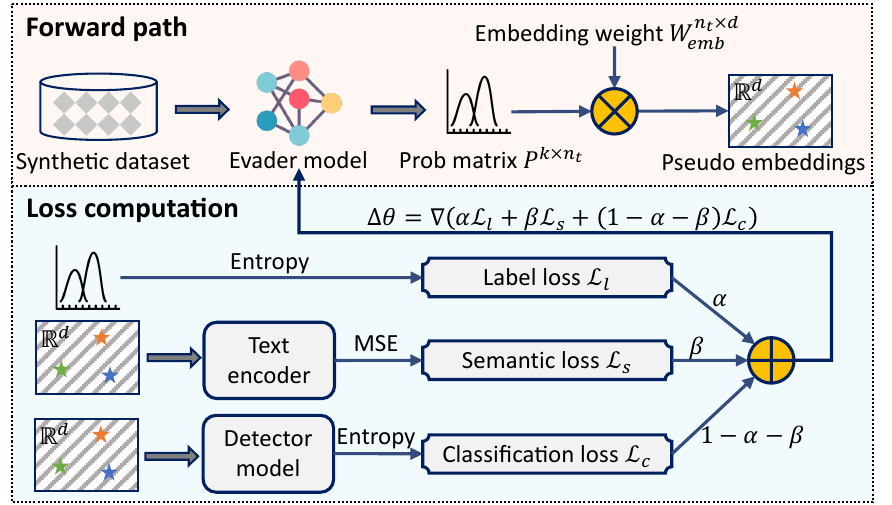}
    \caption{\method training procedure.}
    \label{fig:training_procedure}
\end{figure}

\subsection{Evader Training Details}
\label{sec:white-box_details}

We formulate our \method as an optimization problem and use gradient descent to solve it.
In particular, we transform \autoref{eq:attack_goal} to \autoref{eq:semantic_constraint} into \textit{classification loss}, \textit{label loss}, and \textit{semantic loss}, respectively.
Classification loss serves as the functional constraint, while label loss and semantic loss act as the consistency constraint.
Through classification loss, the evader can acquire new knowledge from the detector model's gradients to overcome the challenge of contribution collapse.
The training procedure of \method is illustrated in \autoref{fig:training_procedure}.

\mypara{Function Constraint}
The purpose of the classification loss is to allow the evader to generate output that spoofs the victim detector.
In the white-box scenario, one can calculate the classification loss by using the output of $\mathcal{F}_\theta$ as the input of $D_V$ through \autoref{eq:attack_goal}.
This method has been widely used in the \textit{compute vision} (CV) domain (e.g., \textit{generative adversarial network} (GAN)) to craft deepfake images.
Nevertheless, it is not commonly used in the NLP domain.
Due to the discrete nature of text, the process of sampling text from the probabilities output by LMs is non-differentiable, hindering the backpropagation of gradients.
To overcome the problem of non-differentiability, we employ a technique named \textit{pseudo embeddings} illustrated in~\autoref{fig:training_procedure}.
Its main idea is to use matrix multiplication to replace the lookup operation in the token embedding layer.
Before passing LMs, a text would first be segmented into tokens and converted into token IDs (see examples in \autoref{tab:tokenization}).
These token IDs are then used to look up corresponding token embeddings in the embedding matrix $W_{emb}^{n_t \times d}$.
Here, $n_t$ is the number of tokens in the LM dictionary, and $d$ is the dimension of token embeddings.
The pseudo-embedding technique replaces token IDs with token probability vectors as inputs, using matrix multiplication to generate their embeddings.
This technique has been proven effective in recent research on text backdoor detection~\cite{azizi2021t, liu2022piccolo}, text backdoor attack~\cite{bagdasaryan2022spinning}, and text adversarial attack~\cite{guo2021gradient}.
Based on pseudo embeddings, our classification loss can be formulated as follows:
\begin{equation}
    \mathcal{L}_c = \mathcal{L}_{ce}\left( D_V\left( P^{k\times n_t} \cdot W_{emb}^{n_t \times d} \right), y_{\text{human}} \right),
\end{equation}
where $\mathcal{L}_{ce}$ stands for cross-entropy loss while $P^{k\times n_t} = \mathcal{F}_\theta \left( \left\{x_i\right\}_{i=1}^n \right)$ is the probability matrix output by the evader.
$k$ is the output sequence length.
$y_{\text{human}}$ represents the label of human-generated texts.
We perform post-processing on $P^{k \times n_{t}}$ to enable the multiplication of $P^{k \times n_{t}}$ and $W_{emb}^{n_{t} \times d}$ in practical implementations.
The key idea of post-processing is to reassign the probability assigned to tokens generated by the evader model but missing from the detector model.
We redistribute this probability to tokens in the detector model.
The details are provided in \autoref{app:post-processing}.

\mypara{Consistency Constraint}
In \textit{reinforcement learning from human feedback} RLHF~\cite{ouyang2022training}, the pairwise KL divergence between the output probability distribution of policy and reference model is used as the consistency constraint.
In \aigt evasion scenarios, our objective slightly differs from RLHF. 
Instead of aiming for the trained model's output to be close to the original model, our goal is for the output of the trained model to be close to the input.
Therefore, we use the distribution between the output and the input as our label loss.
KL divergence is equivalent to cross-entropy when the label is a one-hot probability distribution.
As a result, our label loss is formulated as follows:
\begin{equation}
    \mathcal{L}_l =  \frac{\sum_{i=1}^{n} \mathcal{L}_{ce} \left( P_i, x_i \right)  }{n}.
\end{equation}
One advantage of our label loss is that we do not need to load a reference model on the GPU, which saves half the memory overhead and allows \method to run on consumer-level GPUs.

In addition to label loss, our consistency constraint incorporates a semantic loss.
It is essential because the attacker does not hope the evader will change the text semantics significantly.
A single label loss is not enough, as a few important word changes can completely alter the meaning of a text, such as changing ``yes'' to ``no'', or ``love'' to ``hate''.
We employ \textit{sentence transformers}~\cite{reimers2019sentence} to encode both input and output of $\mathcal{F}_\theta$ into vectors.
Then we use an MSE loss to constrain their $l2$ distance, which is
\begin{equation}
    \mathcal{L}_s = \operatorname{MSE} \left( E_{st}\left( \left\{x_i\right\}_{i=1}^n \right), E_{st}\left( P^{k\times n_t} \cdot W_{emb}^{n_t \times d} \right) \right).
\end{equation}
Here $E_{st}\left( \cdot \right)$ means the input's feature vector obtained by sentence transformers.
We employ a similar pseudo-embedding technique used in classification loss to make the process of computing the output's feature vector differentiable.
In our implementation of \method, $E_{st}\left( \left\{x_i\right\}_{i=1}^n \right)$ is computed prior to training, as the calculation of $E_{st}\left( \left\{x_i\right\}_{i=1}^n \right)$ does not involve gradient propagation.
This pretraining computation reduces training overhead and avoids loading two $E_{st}$ on the GPU.

\begin{table*}[h]
\setlength{\tabcolsep}{4pt}
\centering
\caption{Tokenization examples of LMs on the sentence of ``\textit{Different LMs possess different tokenizers.}''}
\label{tab:tokenization}
\resizebox{\linewidth}{!}{
\begin{tabular}{llll}
\toprule
\textbf{LM} & \textbf{Tokenizer} & \textbf{Tokens} & \textbf{Token IDs} \\
\midrule
BART & BPE & 'Different', 'ĠL', 'Ms', 'Ġpossess', 'Ġdifferent', 'Ġtoken', 'izers', '.' & 44863, 226, 13123, 15256, 430, 19233, 11574, 4 \\
RoBERTa & BPE & 'Different', 'ĠL', 'Ms', 'Ġpossess', 'Ġdifferent', 'Ġtoken', 'izers', '.' & 44863, 226, 13123, 15256, 430, 19233, 11574, 4 \\
GPT2 & BPE & 'Different', 'ĠL', 'Ms', 'Ġpossess', 'Ġdifferent', 'Ġtoken', 'izers', '.' & 40341, 406, 10128, 8588, 1180, 11241, 11341, 13 \\
BERT & WordPiece & 'Different', 'L', '\#\#Ms', 'possess', 'different', 'token', '\#\#izer', '\#\#s', '.' & 14380, 149, 25866, 10192, 1472, 22559, 17260, 1116, 119 \\
\bottomrule
\end{tabular}
}
\vspace{-0.3cm}
\end{table*}

\mypara{Solving Optimization Problem}
After defining the three loss terms $\mathcal{L}_l$, $\mathcal{L}_s$, and $\mathcal{L}_c$, we can formulate our \method to an optimization problem by combine them together:
\begin{equation}
    \min_{\theta} \mathcal{L} = \alpha \cdot \mathcal{L}_l + \beta \cdot \mathcal{L}_s + (1-\alpha - \beta) \cdot \mathcal{L}_c,\label{eq:overall_loss}
\end{equation}
where $\alpha$ and $\beta$ are two hyperparameters to balance the consistency constraint.
A larger $\alpha$ places greater emphasis on syntactic similarity, while a larger $\beta$ emphasizes semantic similarity.
By ensuring that the sum of the coefficients for the tree loss terms equals 1, we can eliminate the impact of the learning rate while tuning $\alpha$ and $\beta$.
Through \autoref{eq:overall_loss}, we can solve the optimization problem using vanilla gradient descent.
During the training process, we freeze the text encoder and the detector model, as they do not require optimization.

\subsection{Warm-started Evader}
\label{sec:tokenizer_unmatch}

Unfortunately, the application of the pseudo embeddings technique encounters the tokenizer mismatch challenge in real-world scenarios, as it implicitly requires the evader and the detector to use the same tokenizer, enabling direct multiplication of $P$ and $W_{emb}$.
However, for CLMs and MLMs, it is not always possible to find corresponding seq2seq models that use the same tokenizer.
Bagdasaryan et al.~\cite{bagdasaryan2022spinning} proposed a \textit{token re-mapping} algorithm to resolve the tokenizer mismatch issue between BART and GPT2.
However, the token re-mapping algorithm can only address the issue of mismatched token IDs, but not the differences in tokenization methods.
As shown in~\autoref{tab:tokenization}, BART and GPT2 differ only in the token IDs.
They both use \textit{byte-pair encoding} (BPE) tokenization method.
The token re-mapping algorithm cannot concatenate BART with BERT, as BERT utilizes \textit{WordPiece} tokenization method.
As far as we know, there is no pretrained seq2seq model that uses the WordPiece tokenizer.

In this paper, we propose a warm-started evader method based on the \textit{warm-started encoder-decoder} technique~\cite{rothe2020leveraging} to enable the construction of evaders against any type of detector LM.
The warm-started encoder-decoder technique is a way to initialize an encoder-decoder model with pretrained encoder and/or decoder-only checkpoints (e.g., BERT, GPT2) to skip the costly pretraining.
It initializes the parameters shared between the encoder-decoder model and the standalone encoder/decoder model using the standalone model's parameters.
Parameters unique to the encoder-decoder model will be randomly initialized.
In essence, this technique provides a mapping from the parameters of stand-alone encoder/decoder models to the parameters of a seq2seq model:
\begin{equation}
    f_{ws}: \left( \theta_{enc}/\theta_{dec}, \theta_{enc}/\theta_{dec} \right) \mapsto \theta_{s2s}.
\end{equation}
We first initialize a seq2seq model using the detector LM and train the seq2seq model to a repeater by SFT.
Specifically, our warm-started evader method includes three steps:
\begin{enumerate}[label=\textbf{Step$\arabic*$:},itemindent=0.4cm, itemsep=0mm]
  \item \textit{Model Initialization.}
  Use function $f_{ws}$ and pretrained detector model to initialize the evader model $\theta_{eva}$.
  \item \textit{Data Collection.}
  Collect public corpora, such as WikiText~\cite{merity2016pointer} and BookCorpus~\cite{Zhu_2015_ICCV}, to form a repeater dataset.
  \item \textit{Repeater Training.}
  Sample text from repeater dataset and update $\theta_{env}$ by SFT.
  The output label is a copy of the input so that the model can repeat the input.
\end{enumerate}
In our experiments of BERT, repeater training can be finished within 5,000 steps with a batch size of 16.
After getting the warm-started evader, the attacker can carry out \method training following~\autoref{sec:white-box_details}.

\section{Opaque Model Attack}
\label{sec:gray-box}

In this section, we elaborate on how \method handles opaque model scenarios, where the attacker only has query access to the victim detector.
Our key idea is to train a surrogate detector to mimic the victim detector.
The attacker first conducts a tokenizer inference attack to get the model architecture, then uses a model extraction attack to train the surrogate detector.
After that, the attacker can train \method against the surrogate detector according to~\autoref{sec:methodology}.

\mypara{Tokenizer Inference Attack}
This attack exploits the fact that current LMs are bound with specific tokenizers.
Although GPT2 and RoBERTa both employ the BPE tokenizer, their special tokens differ.
Specifically, GPT2 tokenizer lacks {\fontfamily{qcr}\selectfont<pad>} and {\fontfamily{qcr}\selectfont<unk>}, which exist in RoBERTa tokenizer.
This discrepancy offers attackers a method to deduce the victim detector's tokenizer and infer the model architecture.
The main thought of our tokenizer inference attack is to query the detector using inputs that are tokenized identically by a particular tokenizer but differently by others.
The attacker can then ascertain if the victim's detector uses the particular tokenizer by comparing the returned scores.
Here we provide two methods to craft inference inputs:
\begin{compactitem}[$\bullet$]
    \item \textit{Padding Spoof.}
    The attacker can add pad tokens to the end of the input text.
    LM tokenizers usually pad the text to ensure inputs have equal length.
    The LM will ignore correct pad tokens, otherwise, it will be tokenized to other tokens.
    For instance, RoBERT pad token {\fontfamily{qcr}\selectfont<pad>} would be tokenized as [``<'', ``pad'', ``>''] by GPT2 tokenizer and BERT tokenizer.
    \item \textit{Space Spoof.}
    The attacker can replace spaces with LM-specific tokens in the input text and observe the response scores before and after replacement.
    Different LM tokenizers handle spaces differently.
    BERT ignores spaces, while LLaMA2 denotes spaces by ``▁''.
    ``Space spoof'' and ``Space▁spoof'' will both be tokenized as [``▁Space'', ``▁spo'', ``of''] by LLaMA2 tokenizer, but would be respectively tokenized as [``Space'', ``s'', ``\#\#po'', ``\#\#of''] and [``\sptoken{[UNK]}''] by BERT tokenizer.
\end{compactitem}

\mypara{Model Extraction Attack}
Model extraction attack aims to construct a substitute model that closely approximates the functionalities or the parameters of the target victim model.
It has been proven effective to steal online models by previous literatures~\cite{tramer2016stealing, orekondy2019knockoff, chen2023d}.
As discussed in~\autoref{sec:threat_model}, model extraction attack consists of three steps: query dataset construction, victim model query, and substitute model training.
In this paper, we employ the vanilla label-only model extraction attack, where the victim detector only exposes predicted labels.
We construct our query dataset using texts drawn from the same distribution as the evader's training dataset, which may differ from the distribution of the detector's training dataset.
We ensure that the query dataset does not overlap with either the evader's or the detector's training datasets.
Then, we use this dataset to query the victim detector and utilize its predictions to label data, creating the retraining dataset.
Finally, we use the retraining dataset to fine-tune the pretrained model whose architecture is known from the tokenizer inference attack.
After obtaining the substitute model, the attacker can use it to compute the classification loss, derive gradients, and subsequently update the evader's parameters.
There are also other opportunities for improvement in model extraction attacks.
However, this area is orthogonal to the research presented in this paper, so we only consider the label-only method in our study.
Also, we find that vanilla label-only method can achieve commendable attack performance (see \autoref{sec:eval_opaque_model_attack}).

\section{Evaluation}
\label{sec:evaluation}

In this section, we first examine the efficacy of \method against three types of LM classifiers in open model scenarios.
Second, we demonstrate that \method can maintain its attack capability in opaque model scenarios.
Third, we conduct dataset transfer experiments to test its transferability.
Fourth, we assess the computation and memory costs of \method.
Fifth, we apply \method to two real-world commercial detectors to demonstrate its practicality.

Additionally, several aspects of our evaluation are documented in the appendix due to space limitations:
(1) We assess the ability of \method to compromise advanced detectors, which feature enhanced model architecture or training strategy, in open model scenarios (\autoref{sec:open_advanced_attack}).
(2) Our evaluation also considers \method against these advanced detectors in opaque model scenarios (\autoref{sec:eval_opaque_advanced_detectors}).
(3) We test our \method against statistic-based detectors through model querying (\autoref{sec:attack_statistic}).
(4) We examine the semantic consistency of texts generated by \method using both GPT4 annotation and human evaluation, and compare the results with four baseline evaders (\autoref{app:semantic_consistency}).

\begin{figure*}[!tbp]
    \centering
    \includegraphics[width=\textwidth]{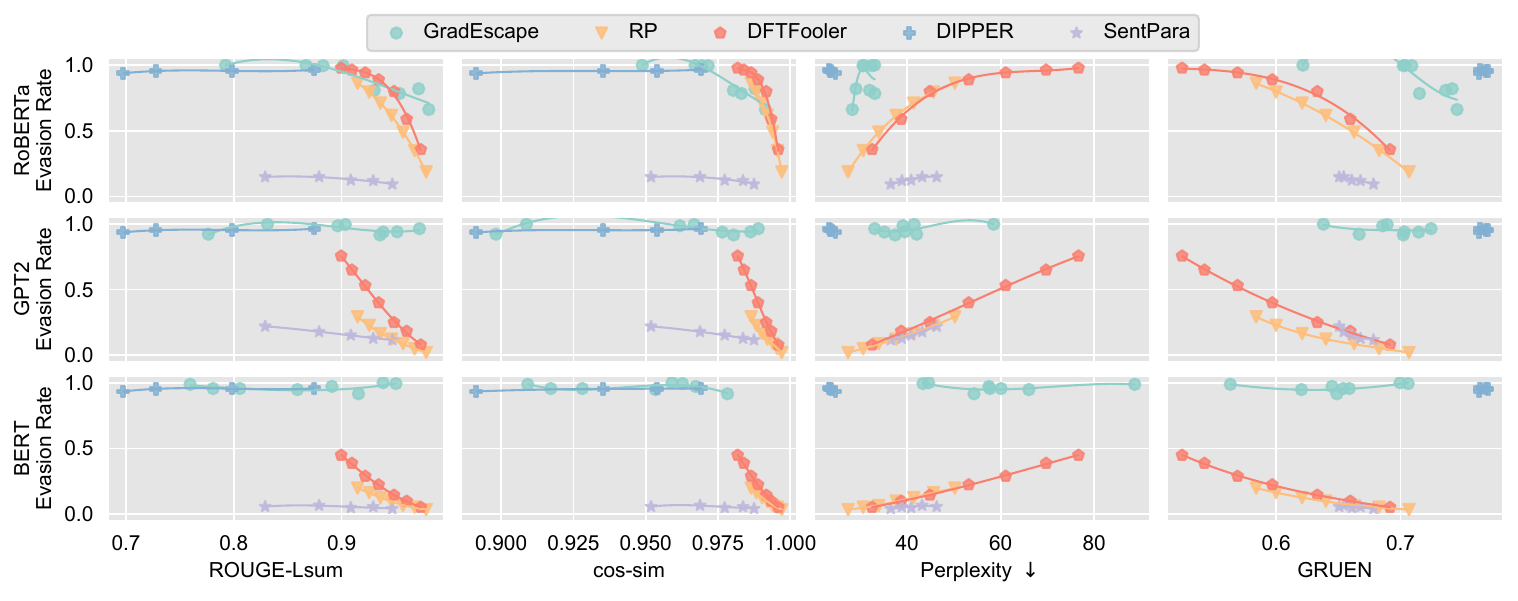}
    \vspace{-0.5cm}
    \caption{Evasion rates versus text quality metrics on GROVER News dataset.}
    \label{fig:white_box_grover}
    \vspace{-0.3cm}
\end{figure*}

\subsection{Experimental Setup}
\label{sec:experimental_setup}

\mypara{Datasets}
We consider four \aigt datasets for evaluation, namely GROVER News~\cite{zellers2019defending}, HC3~\cite{guo2023close}, GPA~\cite{liu2023check}, and GPTWiki Intro~\cite{aaditya_bhat_2023}.
To ensure diversity, we consider both aligned and unaligned LLMs, and consider three tasks: text completion, question-answering, and instruction-writing.
Due to the space limitation, we defer detailed information of datasets to~\autoref{app:detailed_setting}.

Cai et al.~\cite{cai2023evade} found that the differences in format between human-generated text and machine-generated text lead detectors to rely on text format for prediction, thereby reducing the robustness of the detector.
To eliminate the influence of text format, we clean the text before training detectors.
Specifically, we remove line breaks, standardize the use of punctuation, and delete modal particles.

\mypara{Metrics}
We use \textit{evasion rate} (ER) to assess evaders' attack utility.
Given that the attacker aims to maintain syntactic and semantic similarity while ensuring readability, we employ four additional text integrity and quality metrics: \textit{ROUGE}, \textit{cos-sim}, \textit{perplexity}, and \textit{GRUEN}.
ROUGE is for syntactic similarity; cos-sim is for semantic similarity; while perplexity and GRUEN are for readability.
We defer the details of metrics to~\autoref{app:detailed_setting} due to space limitations.

\mypara{Setups of Detectors and Evaders}
In this paper, we select three basic detectors, two advanced detectors, and one statistics-based detector as victim detectors. We also choose four representative evaders as baselines.
The configurations of detectors and evaders largely follow the original papers.
For brevity, we defer the detailed settings in~\autoref{app:detailed_setting}.

\mypara{Implementation}
We use HuggingFace Accelerate~\cite{accelerate} to implement our \method.
Followers can easily run \method using distributed techniques such as Microsoft's DeepSpeed~\cite{rasley2020deepspeed} and Nvidia's Megatron-LM~\cite{shoeybi2019megatron}.
In this work, we use a 2 $\times$ A100-80GB GPU node for all experiments.
Our node possesses 32 CPU cores and 1TB of memory.

\mypara{Disscussion}
Note that our \method does not require the synthetic dataset to share the same distribution as the dataset used to train the detector.
Instead, \method only requires the synthetic dataset to have the same distribution as the texts targeted for evasion.
As discussed in~\autoref{sec:threat_model}, this requirement is easily met since both of them can be generated from the attacker's LLM.
In our evaluations, we primarily focus on attacking detectors with the same dataset distribution as evaders because they are harder to evade.
Otherwise, detectors would suffer from low transferability~\cite{pu2023deepfake}.
\autoref{sec:dataset_transferability} demonstrates that \method is also effective when the data distribution of evaders and detectors does not match.

GradEscape can be easily extended to attack LLM-based detectors.
In this paper, we do not consider fine-tuning LLMs as detectors because we find that LLMs tend to overfit and achieve lower accuracy.
For example, RoBERTa-Base achieves 0.99 accuracy on HC3 and 0.91 accuracy on GROVER, while LLaMA3-8B achieves 0.96 on HC3 and 0.88 on GROVER despite its significant training and inference overhead.

\begin{figure*}[!tbp]
    \centering
    \includegraphics[width=\textwidth]{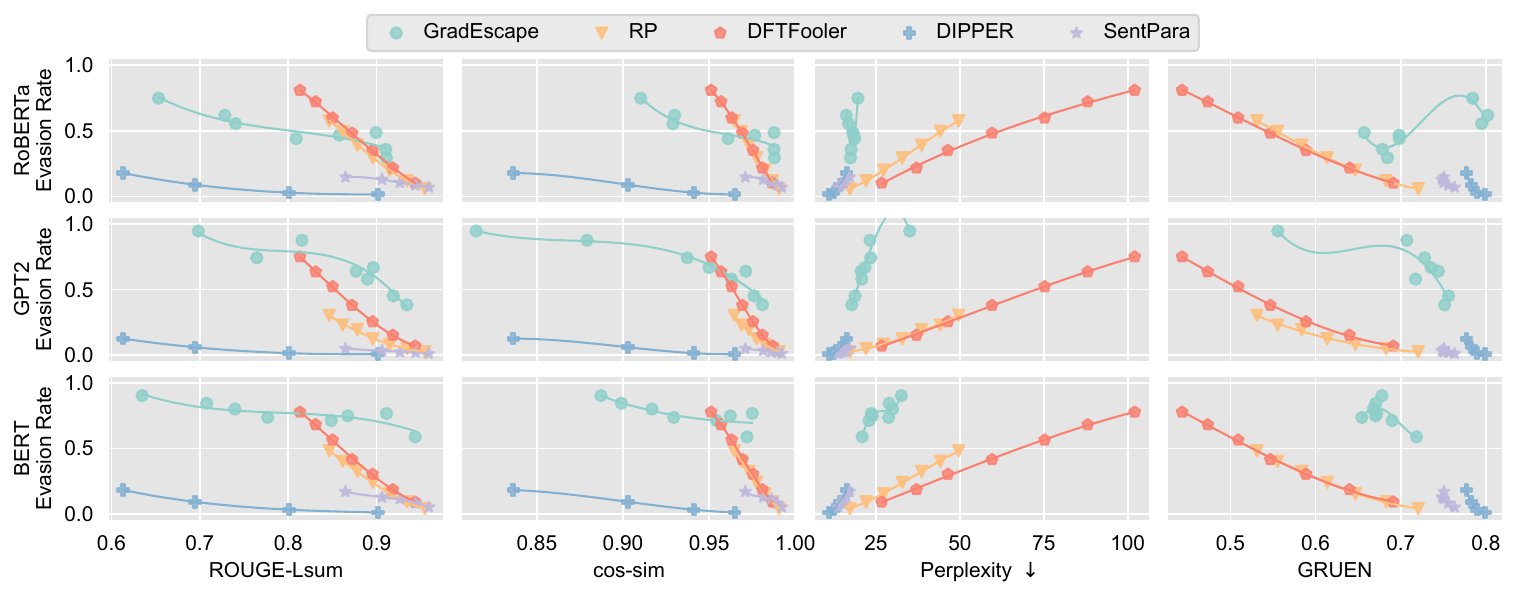}
    \vspace{-0.5cm}
    \caption{Evasion rates versus text quality metrics on HC3 dataset.
    }
    \label{fig:white_box_hc3}
    \vspace{-0.3cm}
\end{figure*}

\subsection{Open Model Attack}
\label{sec:eval_open_model_attack}

\mypara{Attack Setup}
In this section, we select three popular LMs, RoBERTa, GPT2, and BERT, as victim detectors.
The detectors' performance before being attacked is shown in~\autoref{tab:detector_performance} (\autoref{app:extra_white_box}).
All detectors have low baseline evasion rates ($<2\%$), which means nearly all machine-generated texts are correctly classified.
Since BART shares the same tokenizer with RoBERTa, we use BART as the evader model to attack RoBERTa.
We use the token re-mapping technique to re-map BART's logits output for attacking GPT2.
For BERT, we first build a warm-started evader model using the technique described in~\autoref{sec:tokenizer_unmatch}.
We randomly choose 15,000 texts from each of the AlpacaGPT4~\cite{peng2023instruction}, OpenWebText~\cite{Gokaslan2019OpenWeb}, and WikiText~\cite{merity2016pointer} datasets to form the repeater dataset.
Using this repeater dataset, we fine-tune the warm-started encoder-decoder model for 5,000 steps with a batch size of 16.
We choose 9,000 samples for each dataset to train \method.
This evader training set does not overlap with the detector training set.
The detailed evader training settings are shown in~\autoref{tab:hyperparameters}.
After training evaders, we employ beam search and set the number of beams to 4 to generate paraphrased texts.
We compare our \method with 4 state-of-the-art evaders.
We adjust $\alpha$ and $\beta$ to alter the degree of consistency constraint.
For baseline evaders, we adjust the knobs they provide.

\mypara{Results}
To consider text quality while comparing the attack utility of evaders, we plot evasion rate versus text quality metrics in~\autoref{fig:white_box_grover} and~\autoref{fig:white_box_hc3} for GROVER News and HC3, respectively.
Due to space limitations, we defer the results of GPA and GPTWiki to~\autoref{app:extra_white_box}.
For ROUGE, cos-sim, and GRUEN, a position further toward the upper right is preferable, whereas for Perplexity, a position further toward the upper left is preferable.
We can observe that \method is superior to other evaders in most cases.
Although RP and DFTFooler have a higher evasion rate than \method under some ROUGE and cos-sim values, they significantly reduce text readability, indicated by high Perplexity and low GRUEN scores.
We find that DIPPER only achieves a high evasion rate ($>0.8$) on the GROVER dataset, with its evasion rate on other datasets being quite low ($<0.4$).
Note that DIPPER is an 11B model, whereas our \method only has 139M parameters.
\method also achieves a high evasion rate in attacks against BERT detectors, even though we use a warm-started encoder-decoder model as the base model instead of an off-the-shelf seq2seq model.
This demonstrates that our warm-started evader technique can effectively address the issue of tokenizer mismatch.
An interesting finding is that for \method, the evasion rate has a clear linear relationship with ROUGE and cos-sim, but this linear relationship is not significant with Perplexity and GRUEN.
This indicates that the greater the degree of text modification by \method, the easier it is to attack successfully, but the degree of modification does not have a necessary impact on text readability.
We show example outputs of \method in~\autoref{tab:attack_examples} of~\autoref{app:extra_white_box}.
We find that on GROVER News, which is easy to attack, \method only disturbs a few words like a perturbation-based evader, while on more challenging datasets like HC3, \method significantly changes the way of narration like other paraphrase-based evaders.

\begin{figure*}[!tbp]
    \centering
    \includegraphics[width=\textwidth]{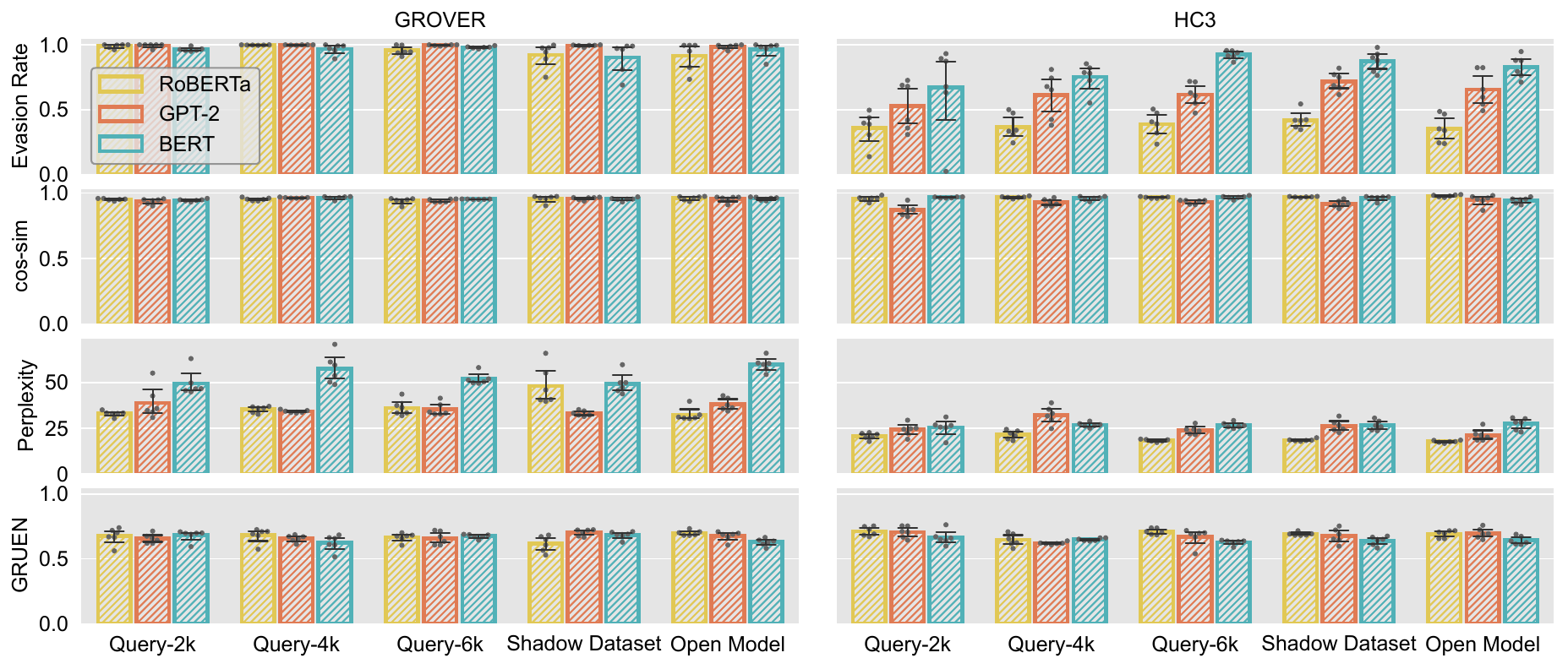}
    \caption{Opaque model attack results.}
    \label{fig:gray_box_results}
    \vspace{-0.3cm}
\end{figure*}

\subsection{Opaque Model Attack}
\label{sec:eval_opaque_model_attack}

\mypara{Attack Setup}
We choose 2,000, 4,000, and 6,000 samples from each dataset to query the victim detector, respectively.
These query data do not overlap with the training data of the victim detectors.
Besides the query attack, we consider a scenario where the attacker does not need to query when the attacker has data from the same distribution as the detector training set to train a surrogate detector.
We choose 12,000 samples from each dataset as the \textit{shadow dataset}.
These shadow datasets also do not overlap with the detector training set.
For each set of experiments, we collect 6 results and adjust $\alpha$ and $\beta$ to ensure the ROUGE scores are between 0.85 and 0.9.
Due to heavy overhead, we only conduct gray-box experiments on GROVER News and HC3.

\mypara{Results}
Our gray-box experimental results are shown in~\autoref{fig:gray_box_results}.
From the figure, we can see that 2,000 queries are sufficient to achieve an evasion rate close to that of a white-box scenario.
Especially on the GROVER News dataset, 2,000 queries can achieve an evasion rate close to 100\%.
The query dataset for 2,000 is only $1/6$ the size of the detector training set.
We find that on the HC3 dataset, the fewer queries there are, the more unstable the evasion rate becomes.
This may be due to the surrogate model being trained with little data, leading to a tendency to overfit.
And evaders trained on an overfitted surrogate model would have poor generalization.
Note that in this paper, we use a vanilla label-only model extraction attack to construct the surrogate model.
Attackers could use improved methods to enhance the generalization of the surrogate model.
Another observation is that query attacks decrease the quality of generated text when attacking RoBERTa and GPT2 detectors.
But this decrease is within 5\%, which is acceptable.
We find that shadow dataset attacks can achieve an evasion rate and text quality close to that of a white-box scenario, which enables \method attack without direct access to the detector.

\subsection{Dataset Transferability}
\label{sec:dataset_transferability}

\begin{figure}[!tbp]
    \centering
    \includegraphics[width=\columnwidth]{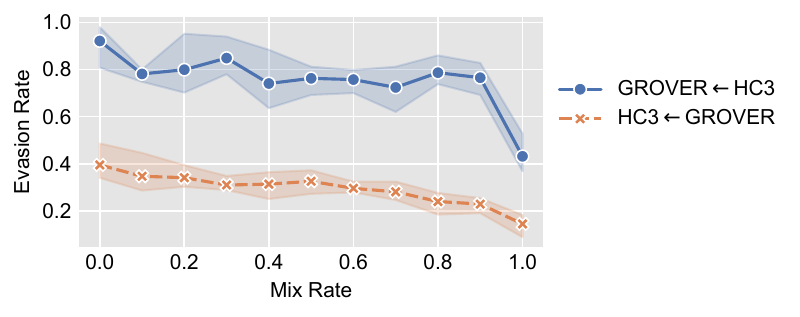}
    \caption{Evasion rates under target AIGT distribution shift. Mix rate is the proportion of external dataset texts mixed into the synthetic training dataset. ``GROVER$\leftarrow$HC3'' denotes training on GROVER mixed with HC3 texts, and vice versa.}
    \label{fig:cross_dataset_results_aigt}
\end{figure}

In this section, we evaluate the capability of \method to generalize across datasets with different distributions. Specifically, we consider two scenarios: (1) the distribution of the evader's training data differs from that of the AIGT targeted for editing; and (2) the distribution of the evader's training data differs from that of the detector's training data.

\mypara{Transfer to Unseen AIGT}
\method relies on a synthetic dataset, composed exclusively of texts generated by LLMs, to train evaders.
As discussed in~\autoref{sec:threat_model}, the attacker can easily construct synthetic datasets by using LLMs that produce target texts intended for editing.
Here, we examine an extreme scenario in which the attacker does not have sufficient AIGT samples that share the same distribution as the target texts to construct a synthetic dataset.
We simulate this situation by mixing texts from external datasets into the synthetic training data while keeping the test set unchanged.
In particular, we choose GROVER and HC3 datasets and mix in texts from the other dataset to train evaders. Then we measure evasion rates against RoBERTa at a fixed ROUGE score around 0.9.

We change the proportion of mixture and report evasion rates in~\autoref{fig:cross_dataset_results_aigt}.
We vectorize the datasets using TF-IDF with an n-gram range of 2--3 and measure the cosine similarity between mixed and original datasets, which gradually decreases from 1.0 to 0.495 as the mix rate increases.
Results show that evasion rates decline with increasing mix rates but remain above half of the original one when the mix rate is below 1.0.
A significant drop occurred only when the mix rate reached 1.0, i.e., training exclusively on external datasets.
However, this situation is unlikely, as attackers can at least include the targeted texts as the evader's training data.

\mypara{Transfer to Heterogeneous Detectors}
In~\autoref{sec:gray-box}, we discuss how attackers can easily obtain the model architecture of the detector; thus, aligning the surrogate model's architecture with that of the victim detector is straightforward.
However, attackers may occasionally find that the data they intend to use for an attack does not align with the training data of the victim detector.
For example, an attacker may wish to use AI to assist in writing a paper, but the victim detector is trained on a question-answering dataset.
Below, we assess \method's effectiveness in such a situation where the distributions of evaders' and detectors' datasets significantly vary.
We choose surrogate model training data and evader training data from the same dataset.
We select RoBERTa, MPU, and GLTR as our victim detectors since they perform best in their respective categories.
For GLTR, we train a logistic regression model using its word-rank features for classification.
Our GLTR implementation is the same as~\cite{pu2023deepfake} and~\cite{he2023mgtbench}.
We adjust $\alpha$ and $\beta$ to achieve a target ROUGE of approximately 0.9 and measure the evader's evasion rate against the victim detector.

\begin{figure}[t]
    \centering
    \includegraphics[width=\columnwidth]{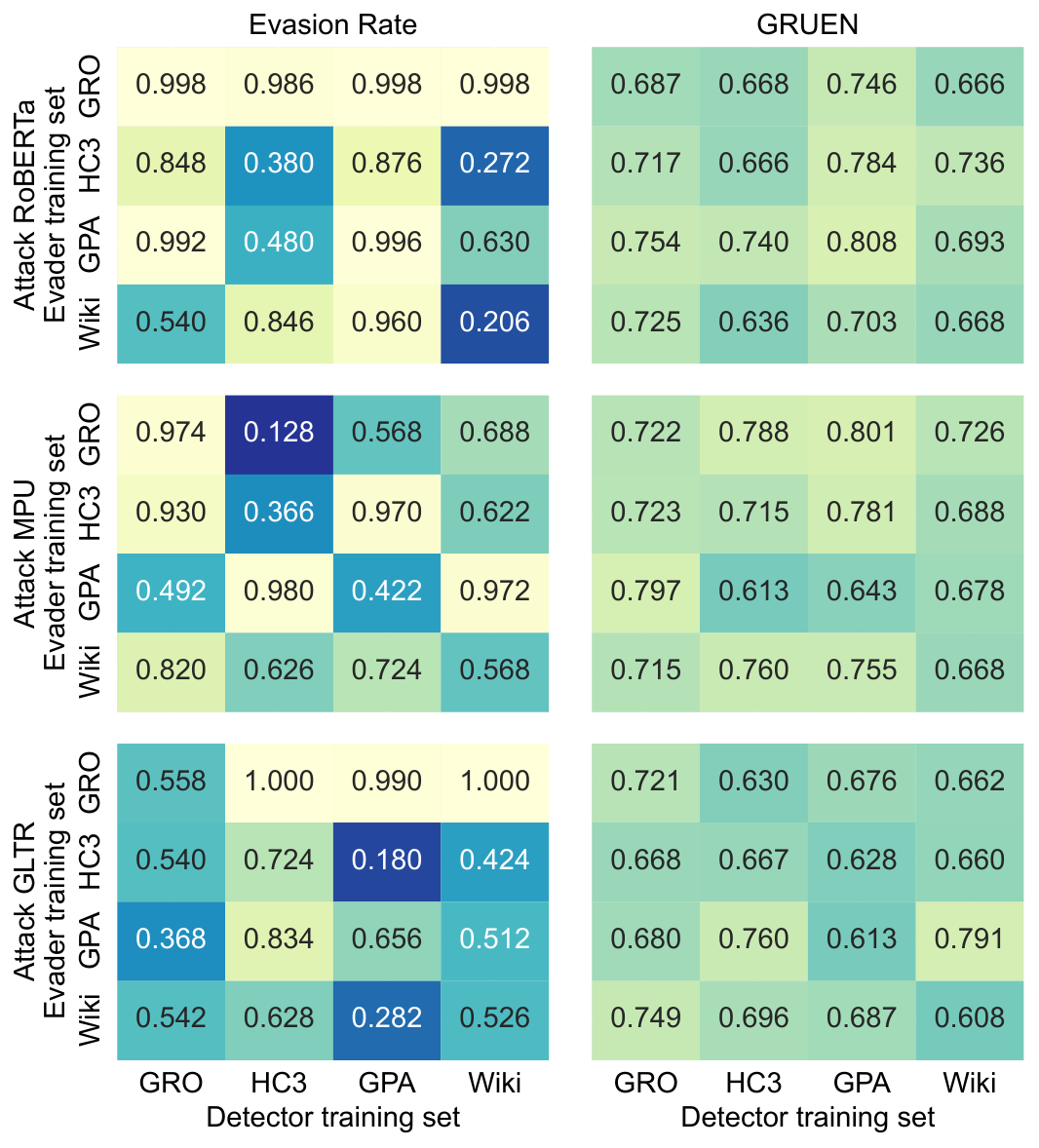}
    \caption{Evasion results under detector distribution shift.}
    \label{fig:cross_dataset_results_detector}
\end{figure}

We report the evasion rates and GRUENs in~\autoref{fig:cross_dataset_results_detector}.
We discover that using different training sets for evader and detector typically results in higher evasion rates without significantly impacting the text quality.
This may be due to the inherently poor data transferability of machine learning models, which leads to the subpar performance of the victim detector when identifying texts that are not from the same distribution as its training set.
Evasion rates are more closely related to the evader's training set than to the victim detector training set.
For instance, GROVER News, a dataset that is easy to attack, achieves nearly 100\% evasion rates across all victim detectors, while HC3, a dataset that is more challenging to attack, has lower evasion rates on all victim detectors.
To further assess \method's transferability against detectors trained on heterogeneous datasets, we compare its performance with four baseline evaders.
Due to the space limitation, we defer the detailed results in~\autoref{app:evade_heterogeneous}.
While distribution discrepancies degrade \method's performance, it still outperforms baselines in many cases.

\begin{table}[tbp]
    \centering
    \caption{Evaders' computation and GPU memory overhead.
    We measure the time required to complete attacking 100 samples and the maximum GPU memory footprint.}
    \label{tab:overhead}
    \resizebox{\columnwidth}{!}{
    \begin{tabular}{l|ccccc}
    \toprule
    Evader & RP & DFTFooler & DIPPER & SentPara & \method \\
    \midrule
    Time & 382s & 1182s & 412s & 329s & 105s \\
    GPU Mem & 3.51GB & 9.54GB & 43.51GB & 4.43GB & 2.48GB \\
    \bottomrule
    \end{tabular}
    }
    \vspace{0.2cm}
\end{table}

\subsection{Cost Analysis}

The overhead of an evader includes computational expenses and GPU memory usage.
We measure the overhead for \method and other baseline evaders.
When measuring DIPPER, SentPara, and \method, we set the batch size to 1 and use only one GPU.
The results are shown in~\autoref{tab:overhead}.
\method has the smallest GPU memory footprint and the shortest computation time among all evaders.
While DIPPER boasts considerable attack utility, its inference requires more than 40GB of GPU memory, rendering it impractical for consumer-grade graphics cards.
Our \method offers the advantages of a smaller model size and lower training costs.
In our experiments, with a batch size of 64 and set for 5 epochs, \method could complete its training within 50 minutes.
Due to \method's minimal space requirements, attackers could potentially distribute the evader to sub-attackers for use.

\begin{table*}[tbp]
    \centering
    \caption{Evasion results of real-world case studies.}
    \label{tab:real_world}
    \resizebox{\textwidth}{!}{
    \begin{tabular}{llcccccccccc}
    \toprule
    & & ROUGE & cos-sim 
    & \multicolumn{2}{c}{PPL$\downarrow$} & \multicolumn{2}{c}{GRUEN} 
    & \multicolumn{2}{c}{$\overline{\text{DC}}\downarrow$} & \multicolumn{2}{c}{ER} \\
    \cmidrule(lr){5-6} \cmidrule(lr){7-8}
    \cmidrule(lr){9-10} \cmidrule(lr){11-12}
    Detector & \diagbox{Dataset}{ATK} & \checkmark & \checkmark
    & \textbf{--} & \checkmark & \textbf{--} & \checkmark
    & \textbf{--} & \checkmark & \textbf{--} & \checkmark \\
    \midrule
    \multirow{2}{*}{Sapling} & HC3 & 0.852 & 0.965 & 9.2 & 25.7(+16.5) & 0.801 & 0.685(-0.116) & 0.999 & 0.550(\textbf{-0.449}) & 0.000 & 0.448(\textbf{+0.448}) \\
    & GPTWiki & 0.861 & 0.960 & 12.8 & 27.7(+14.9) & 0.758 & 0.670(-0.088) & 0.965 & 0.422(\textbf{-0.543}) & 0.026 & 0.578(\textbf{+0.552}) \\
    \midrule
    \multirow{2}{*}{Scribbr} & HC3 & 0.870 & 0.965 & 9.2 & 19.7(+10.5) & 0.801 & 0.749(-0.052) & 0.982 & 0.378(\textbf{-0.604}) & 0.018 & 0.785(\textbf{+0.767}) \\
    & GPA & 0.884 & 0.913 & 13.7 & 28.5(+14.8) & 0.829 & 0.691(-0.138) & 0.992 & 0.496(\textbf{-0.496}) & 0.008 & 0.658(\textbf{+0.650}) \\
    \bottomrule
    \end{tabular}
    }
    \vspace{-0.3cm}
\end{table*}

\subsection{Real-world Case Studies}
\label{sec:real_world_case_studies}

In this section, we apply \method to two real-world \aigt detection services provided by Sapling~\cite{sapling} and Scribbr~\cite{sapling} to showcase the vulnerability of existing AI-text detection methods.
We first assess the performance of these two detectors across our four datasets in the absence of attacks.
The results are presented in~\autoref{tab:test_real_world}, located in~\autoref{app:extra_real_world}.
We observed that the performance of the two real-world detectors varies across different datasets.
For instance, their detection accuracy on HC3 is higher than 0.91, but they exhibit low recall on GROVER News, meaning they struggle to identify GROVER-generated texts.
For each detector, we select two datasets where the performance is comparatively better for conducting attacks.
We also compare \method with four baselines against above two real-world detectors.
The experimental results show that \method achieves state-of-the-art evaion rates, and Sapling is more robust than Scribbr when confronting evaders.
Due to the space limitation, we defer the comparison to~\autoref{app:extra_real_world}.

\mypara{Attack Sapling}
Sapling offers \aigt detection service through an HTTP API and a free online demo.
Both the HTTP API and the online demo provide the probability that a given text is fake, which we refer to as detection confidence (DC) in this paper. When DC exceeds 0.5, we classify the text as AI-generated.
Initially, we leverage its online demo to execute a tokenizer inference attack, aiming to deduce the architecture of its detector.
We begin by creating a text with a medium DC, which can be easily done by merging a segment of human-generated text with a piece of \aigt.
Following this, we conduct insertion tests and observe the DC variation.
We get the following observations: a) Inserting spaces before punctuations, DC changes; b) Appending \sptoken{<pad>}s after the text, DC does not change; c) Prepending \sptoken{<|endoftext|>}s to the text, DC changes.
These observations are illustrated in~\autoref{fig:tokenizer_inference_sapling} of~\autoref{app:extra_real_world}.
From the above three observations, we can conclude three key points: (1) Sapling detector's tokenizer is sensitive to spaces; (2) Its tokenizer employs \sptoken{<pad>} as the padding token; (3) Its tokenizer does not utilize \sptoken{<endoftext>} as the padding token.
Among the three detector LM architectures considered in this paper (i.e., BERT, RoBERTa, and GPT2), RoBERTa is the only one that meets all the aforementioned criteria.
Though we cannot assert Sapling detector is definitively trained on a RoBERTa model, RoBERTa should be appropriate as Sapling detector's surrogate model since they share similar tokenizers.

\begin{figure}[tp]
    \centering
    \resizebox{\columnwidth}{!}{
    \includegraphics{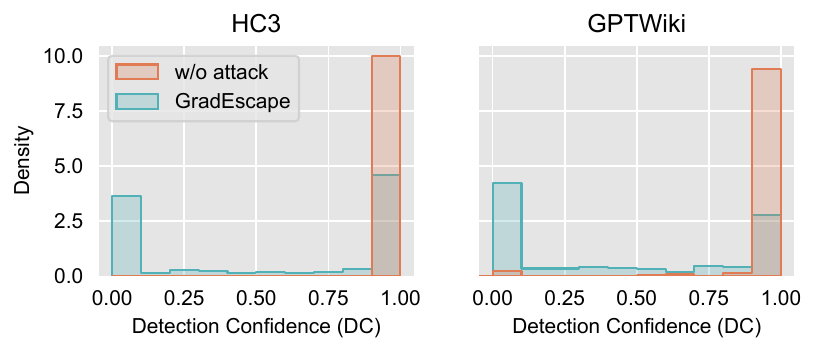}
    }
    \caption{DC histogram of Sapling.}
    \label{fig:sapling_distributions}
    \vspace{-0.3cm}
\end{figure}

\begin{figure}[tp]
    \centering
    \resizebox{\columnwidth}{!}{
    \includegraphics{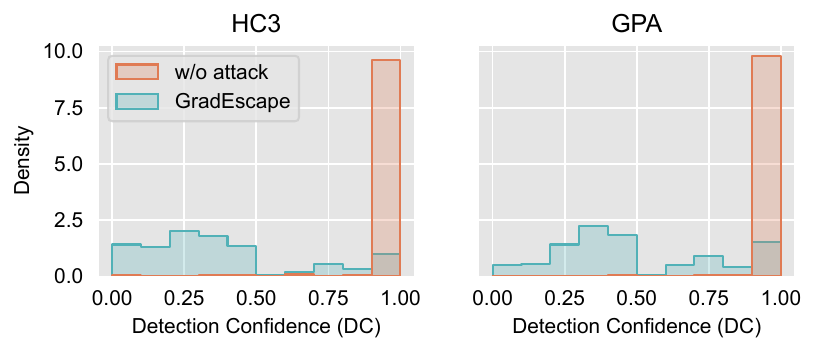}
    }
    \caption{DC histogram of Scribbr.}
    \label{fig:scribbr_distributions}
\end{figure}

After determining the appropriate surrogate model architecture, we construct our retraining dataset.
Given Sapling detector's minimum input requirement of 50 words, we filter out examples shorter than 50 words from the training sets of HC3 and GPTWiki, then randomly select 2,000 samples to form our query datasets.
We utilize Sapling's HTTP API to conduct the querying process.
The querying costs for HC3 and GPTWiki are \$10.10 and \$10.15, respectively.
We retain only the labels obtained from the queries to build our retraining dataset, on which we fine-tune the surrogate model.
Subsequently, we train an evader on this surrogate model, maintaining ROUGE scores within the range of 0.85 to 0.90.
The attack results are depicted in~\autoref{tab:real_world}, and the distributions of DC are presented in~\autoref{fig:sapling_distributions}.
Our results demonstrate that, while slightly compromising text quality, \method is able to reduce the DC of half the texts from around 1 to near 0, achieving an evasion rate of approximately 0.5.
We present an attack example in~\autoref{fig:sapling_attack_example}.

\mypara{Attack Scribbr}
Scribbr also offers a free online demo that outputs DC values.
Following the same steps with Sapling, we conduct a tokenizer inference attack on the Scribbr online demo.
However, we find that all three types of insertions affect the DC, unlike Sapling.
At this point, we are unable to identify a tokenizer that matches Scribbr's tokenization method.
We suspect this is due to Scribbr employing a statistic-based detector, or possibly using a combination of statistic-based and deep learning-based detectors for its assessments.
In~\autoref{sec:attack_statistic}, we show that \method is also effective against statistic-based detectors.
Here, we still choose RoBERTa as the surrogate model despite the inconsistency of tokenization.

Similar to Sapling, Scribbr imposes a restriction on the input length, requiring between 25 to 500 words.
We filter the HC3 and GPA training sets to exclude examples outside this range and then choose 2,000 examples to form our query dataset.
Unlike Sapling, Scribbr does not offer an API.
To implement the querying process, we employ Selenium~\cite{selenium} to write browser automation scripts, assisting us in collecting DC values output by the online demo.
Subsequently, we train the surrogate model and evader using the same approach as with Sapling.
The attack results are shown in~\autoref{tab:real_world}, with the distribution of DC illustrated in~\autoref{fig:scribbr_distributions}.
We find that after the attack, the DC distribution for Scribbr is more uniform.
\method achieves an evasion rate of over 0.6 on both datasets, which indicates \method is still effective when the detector architecture is not assured.
An example of the attack is depicted in~\autoref{fig:scribbr_attack_example} of~\autoref{app:extra_real_world}.

\section{Potential Defense}
\label{sec:potential_defense}

From the aforementioned attack experiments, we can know that deep learning-based methods are susceptible to subtle variations in textual expression.
An evader can modify the text's expression to bypass detection while preserving the underlying semantics.
Several text adversarial attacks~\cite{jin2020bert, li2019textbugger} exploit this vulnerability by substituting essential words with their synonyms, thereby inducing NLP models to generate incorrect outputs.
Ideally, detectors should evaluate texts based solely on semantics. 
However, in our training dataset, there is a consistent difference in expression between AIGT and human-generated text.
For instance, human-generated text tends to be more colloquial and may include spelling errors.
This discrepancy makes it easier for detectors to recognize and use these differences in expression as criteria for detection.

\begin{figure}
    \centering
    \includegraphics[width=0.95\columnwidth]{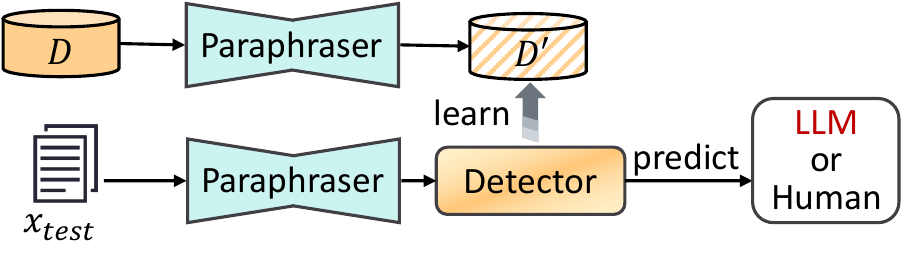}
    \caption{Active paraphrase.}
    \label{fig:paraphrase_defense}
\end{figure}

\begin{table}[tbp]
    \centering
    \caption{Comparison of detection utility between vanilla detector (VD) and paraphrase detector (PD).}
    \label{tab:pd_utility}
    \setlength{\tabcolsep}{3pt}
    \resizebox{\columnwidth}{!}{
    \begin{tabular}{lcccccc}
    \toprule
    & \multicolumn{2}{c}{Accuray} & \multicolumn{2}{c}{FPR} & \multicolumn{2}{c}{$\text{ER}_\text{b}$(FNR)} \\
    \cmidrule(r){2-3} \cmidrule(r){4-5} \cmidrule(r){6-7}
    & VD & PD & VD & PD & VD & PD \\
    \midrule
    GROVER & 0.917 & 0.609(\textbf{-0.308}) & 0.148 & 0.643(\textbf{+0.495}) & 0.016 & 0.131(\textbf{+0.115}) \\
    HC3 & 0.995 & 0.944(\textbf{-0.051}) & 0.006 & 0.100(\textbf{+0.094}) & 0.004 & 0.010(\textbf{+0.006}) \\
    GPA & 0.994 & 0.981(\textbf{-0.013}) & 0.012 & 0.018(\textbf{+0.008}) & 0.000 & 0.019(\textbf{+0.019}) \\
    GPTWiki & 0.983 & 0.942(\textbf{-0.041}) & 0.034 & 0.104(\textbf{+0.070}) & 0.000 & 0.010(\textbf{+0.010}) \\
    \bottomrule
    \end{tabular}
    }
    \vspace{0.2cm}
\end{table}

\mypara{Proposed Defense}
We propose employing active paraphrasing prior to building detectors, as illustrated in~\autoref{fig:paraphrase_defense}.
For a training dataset $D$, we utilize a paraphraser to rephrase each $x \in D$ to obtain $x'$.
Along with the label $y$, this forms the new dataset $D'$.
We then train our detector using $D'$.
During the inference stage, any test text $x_{test}$ is first processed by the paraphraser and subsequently by the detector. 
This paraphrasing step helps minimize discrepancies in text expression during both training and testing phases, thereby allowing the detector to focus exclusively on semantics.
We use Llama-3-8B-Instruct~\cite{llama3modelcard} as our paraphrase, and set the user prompt ``Rewrite the following text for me:\textbackslash n\textbackslash n\{\textit{input text}\}''.
Our defense strategy can be integrated with any post-hoc detectors, as it only involves modifications to the data preprocessing stage.
Notably, our defense remains robust in open model attack scenarios, where the attacker has complete knowledge of both the detector and the paraphraser.
Under such conditions, the attacker is unable to train a gradient-based evader because the detector and paraphrase could employ different tokenizers, so the pseudo-embedding technique is not applicable.
In our \method, we address this challenge by warm-starting an encoder-decoder model as a repeater.
However, there is no technique available to initialize a paraphrase model without considerable overhead.

\begin{figure*}
    \centering
    \includegraphics[width=\textwidth]{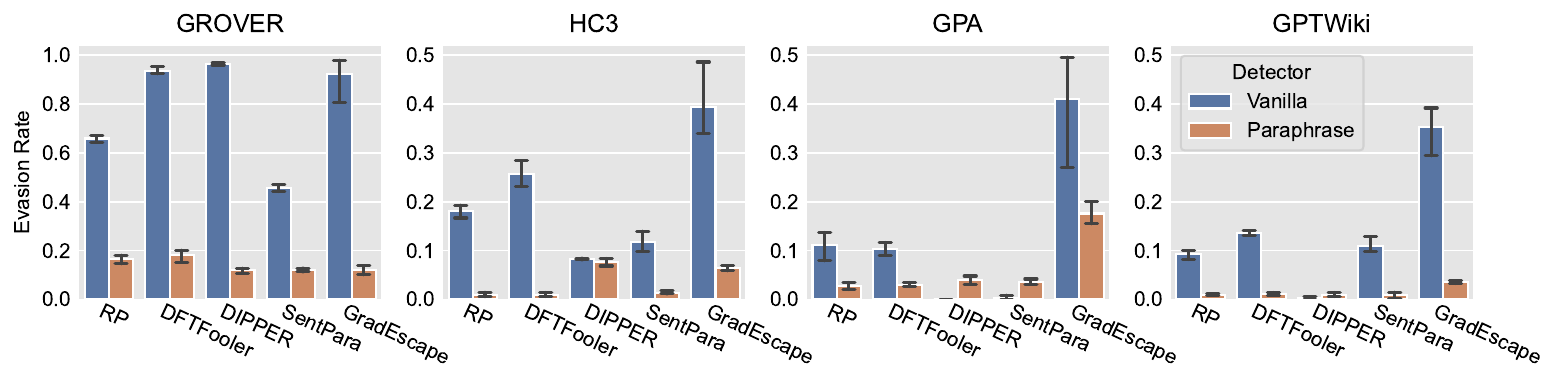}
    \caption{Defense effectiveness of our active paraphrase.}
    \label{fig:pd_defense_effectiveness}
    \vspace{-0.2cm}
\end{figure*}

\begin{figure}
    \centering
    \includegraphics[width=\columnwidth]{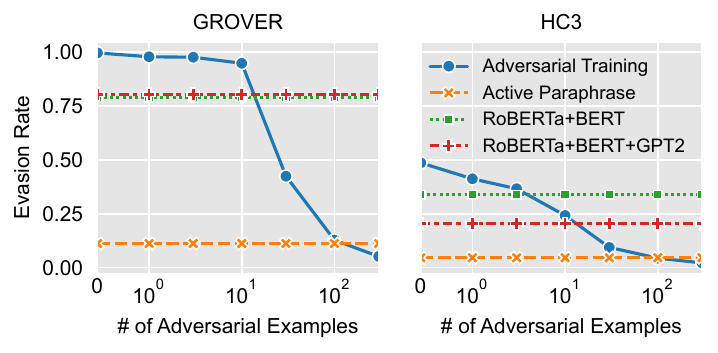}
    \caption{Comparison of active paraphrasing, adversarial training, and ensemble detectors.}
    \label{fig:adaptive_defense}
    \vspace{0.2cm}
\end{figure}

\mypara{Evaluation}
We employ RoBERTa to build our detectors and initially assess the impact of active paraphrasing on detection utility.
The experimental results are presented in~\autoref{tab:pd_utility}.
We observe that, except for GROVER dataset, the accuracy decrease induced by active paraphrasing in the other three datasets remains below 5\%, and the evasion rate increases by less than 2\%.
Active paraphrasing significantly impacts the detection utility on GROVER dataset, possibly due to the smaller semantic differences between AI-generated and human-generated text. This also explains the susceptibility of the GROVER dataset to attack, as revealed in~\autoref{sec:eval_open_model_attack} and~\autoref{sec:eval_opaque_model_attack}.
Experimental results on defense utility are shown in~\autoref{fig:pd_defense_effectiveness}.
Active paraphrasing effectively reduces the evasion rate to below 20\% on GROVER dataset and below 10\% on other datasets, with the only exception of dealing with our \method on GPA dataset.
While active defenses can mitigate most attacks, employing LLM-based paraphrasing for training sets and inputs incurs substantial computational and time overheads, along with a subtle reduction in detection utility.
Whether to use this defensive approach requires careful consideration by defenders.

\mypara{Comparison with Adaptive Defenses}
We compare our proposed defense with two adaptive defenses, assuming that the detector developer is aware of \method.
The first adaptive defense is \textit{adversarial training}~\cite{goodfellow2014explaining}: the detector developer collects texts generated by the evader, augments the original training dataset with these texts labeled as LLM-generated, and then retrains the detector on the augmented dataset.
The second adaptive defense is \textit{ensemble detectors}~\cite{krizhevsky2012imagenet}.
In this approach, the developer trains multiple detectors using different models, and the final detection result is obtained by aggregating the outputs from each model.
We apply soft voting, which averages the predicted probabilities of all models, as the ensemble method.
For evader training, we construct substitute detectors using RoBERTa by querying the ensemble detectors.
The experimental results are shown in~\autoref{fig:adaptive_defense}.
As more adversarial examples are added, adversarial training significantly lowers evasion rates, but it requires over 100 examples to outperform active paraphrasing. Because adversarial training is reactive, the developer must first suffer some attacks, while active paraphrasing provides proactive defense.
We also find that adversarial training does not transfer well between different evaders.
For example, on the GROVER dataset, 200 adversarial examples from one \method evader reduce the evasion rate to 0.110, but using a different \method evader increases it back to 0.712.
This is due to the randomness in GradEscape’s training.
Ensemble detectors are less effective; with three detectors, the evasion rate only drops to half of the original on HC3.
This suggests that different models trained on the same data may share similar weaknesses.

\section{Related Work}
\label{sec:related_work}

\mypara{Text Revision}
Our work can be regarded as a special form of text revision.
The purpose of text revision is to alter certain properties of a text without changing its semantic meaning.
Its applications include text detoxification~\cite{hallinan2023detoxifying, dehghan2022grs}, grammatical error correction~\cite{mita2022towards, dwivedi2022editeval}, text style transfer~\cite{yang2018unsupervised, he2019probabilistic}, and even code editing~\cite{chakraborty2020codit}.
In \method, the property being altered ensures the text evades the target \aigt detector.
Most text revision methods rely on supervised learning and require parallel datasets, which are difficult for evasion attackers to obtain.
Existing non-parallel text revision methods are limited by their performance~\cite{du2022understanding} or the need for a large target corpus~\cite{liu2021non}.
We address the above challenge through unsupervised learning, where the evader's parameters are updated based on the gradients provided by the detectors.

\mypara{Textual Adversarial Attack}
Our work can also be deemed as a textual adversarial attack against detector models.
Textual adversarial attacks aim to induce the victim model to produce erroneous output (e.g., misclassification) by perturbing the original text.
According to a recent study in this field~\cite{zeng2021openattack}, textual adversarial attacks can be categorized into three categories in terms of attacker accessibility, which are blind, decision-based, score-based, and gradient-based.
BadCharacters~\cite{boucher2022bad}, which replaces characters with their homoglyphs, is a blind attack.
SEAR~\cite{ribeiro2018semantically} leverages a paraphrasing model to generate adversarial examples that help debug models by extracting rules.
TextFooler~\cite{jin2020bert} employs a score-based strategy to identify keywords and replace them effectively.
On the other hand, TextBugger~\cite{li2019textbugger} also relies on returned scores for selecting replacements but discerns important words using gradients.
GBDA~\cite{guo2021gradient} is the first pure gradient-based adversarial attack in the NLP domain, utilizing gradients to pinpoint both crucial words and their alternatives.
Our \method can also be categorized as a gradient-based adversarial attack.
However, distinctively, \method eliminates the necessity for attackers to have access to the target or surrogate model after finishing evader training.
Thus, an attacker can distribute \method evader, and anyone who acquires the evader can conduct attacks in a plug-and-play manner.

\mypara{LM Alignment}
Our attack is analogous to LM alignment, but it aligns with detector predictions rather than human preferences.
LM alignment techniques can be categorized into online and offline training, depending on the need for interaction between policy and reward.
The most well-known online training method is RLHF~\cite{ouyang2022training}, which initially conducts SFt and then optimizes the SFT model using PPO algorithm~\cite{schulman2017proximal} based on the feedback from a reward model.
Some work uses language model feedback (RLAIF) as an alternative to human feedback~\cite{bai2022constitutional, lee2023rlaif, pang2023language}.
However, online training often faces optimization instability~\cite{rafailov2024direct, wu2023pairwise} and sensitivity to hyperparameters~\cite{zheng2023secrets, engstrom2020implementation}.
Some researchers have proposed using SFT to replace the RL process~\cite{rafailov2024direct, ethayarajh2024kto, hong2024reference}.
DPO~\cite{rafailov2024direct} transforms the reward modeling stage into a preference learning stage, while DRPO~\cite{hong2024reference} introduces an odds ratio penalty to the SFT loss.
These methods typically employ contrastive pairwise data.
POR~\cite{song2024preference} and RRHF~\cite{yuan2023rrhf} improve preference learning with ranked data.

\section{Conclusion}
\label{sec:conclusion}

We propose a novel gradient-based \aigt evader, named \method, to paraphrase AI-generated text and then bypass post-hoc \aigt detectors.
By efficiently exploiting vulnerabilities in \aigt detectors, \method achieves high evasion rates with a small model size.
We develop a warm-started evader technique, enabling \method to attack detectors of any LM architecture.
Through tokenizer inference attack and model extraction attack, we successfully extend \method into gray-box scenarios.
Extensive experiment results have confirmed that \method's attack utility surpasses that of state-of-the-art \aigt evaders while ensuring text quality.
To mitigate the threat of \method, we further propose a new black-box defense that eliminates potential harmful modifications in the input text.

\section{Acknowledgements}

We thank our anonymous reviewers for their valuable feedback.
This research received support from the National Natural Science Foundation of China under Grant No. 62302441, 62441618. This project was also supported by Open Fund of Anhui Province Key Laboratory of Cyberspace Security Situation Awareness and Evaluation, Ant Group, and the Key Research and Development Program Project of Ningbo Grant No. 2025Z029.

\section{Open Science}
In accordance with the open science policy, we publicly release all artifacts necessary to reproduce the results in this paper, including our experimental datasets, source code, and four evader models targeting real-world AIGT detectors.
Although Scribbr has updated its service to employ stronger detectors, we provide a webarchive file so that readers can access the previous version of Scribbr's service and replicate our experiments about Scribbr.
All stages of our experimentation are available, including data preprocessing, detector training, evader training, and metric computation.
Additionally, we provide a Python library named \texttt{AIGT} to streamline the replication of existing AIGT detectors, thereby promoting transparency and reproducibility in AI-generated text detection research.
Our code, datasets, and models can be accessed through \url{https://doi.org/10.5281/zenodo.15586856}.

\section{Ethical Considerations}

Our research aims to expose the vulnerabilities in current AIGT detectors, assisting developers in creating more robust systems. We weighed the potential benefits and risks of this research and concluded our strategies can effectively mitigate such risks.

\mypara{Disclosure and Potential Misuse}
Similar to other adversarial attack studies \cite{boucher2022bad,slocum2024doesn,fang2024zero}, the newly discovered vulnerabilities in our work may provide insight for potential attackers.
Nonetheless, we believe that responsible disclosure is generally more beneficial than withholding such findings. 
To prevent potential misuse, our team has established strict internal regulations and requirements.
All code of GradEscape was developed and maintained solely by the first author, who is the only person with access to it. We stored our trained evaders on a server that requires public key authentication to log in, lest they be stolen by unauthorized parties. No individuals outside our team have access to the server.

\mypara{Possible Defense}
We propose an active paraphrasing defense method to neutralize the adversarial modifications made by evaders.
Experimental results show that this approach can reduce the evasion rate to below 0.2 at the cost of additional computational overhead for paraphrasing.
For overhead-sensitive scenarios, developers may opt for a perplexity-based filter, as our findings suggest that strong attacks often increase text perplexity. 
However, perplexity filters can cause a lot of false positives when the input is from a diverse source \cite{liang2023gpt}.
A recent study introduced certified robustness techniques into NLP classification models \cite{zhang2024text}, offering rigorous guarantees against textual adversarial attacks.

\mypara{Responsible Real-world Deployment and Human Evaluation}
We implement several measures to ensure our real-world studies do not harm actual systems or users. For the dataset, we exclusively choose publicly available corpora.
The human-generated texts were sourced from open-access websites such as Reddit, Wikipedia, and OpenWebText.
Querying cloud-based AIGT detectors with these texts does not involve private data.
When querying Sapling, we employed the official API and paid a total of {\$}20.25 for building evaders.
For Scribbr, we employed an automated script to simulate user interactions with the WebUI.
To avoid placing an unnecessary load on Scribbr’s infrastructure, all queries were sent at night in U.S. Eastern Time.
They have updated their service models.
Therefore, publishing our work poses no severe harm to these platforms.
For human evaluation, we consulted and obtained approval from our IRB and paid our crowdworkers at a rate exceeding the minimum hourly wage in our hiring region.

\bibliographystyle{IEEEtran}
\bibliography{main}

\section*{Appendix}

\appendix

\section{Detailed Experimental Setup}
\label{app:detailed_setting}

\mypara{Datasets}
We use the following four \aigt datasets for evaluation.
To ensure diversity, we consider both RLHF-trained and non-RLHF-trained LLMs.
Additionally, we consider three tasks: text completion, question-answer, and instruct-writing.

\begin{table*}[tbp]
\centering
\caption{Hyperparameters setting of detectors and \method.}
\label{tab:hyperparameters}
\resizebox{\textwidth}{!}{
\begin{tabular}{ll|ccccccccc}
    \toprule
    & \textbf{Model} & \textbf{\#. Human Texts} & 
    \textbf{\#. LLM Texts} & \textbf{\#. Epoch} & \textbf{Batch Size} & \textbf{Learning Rate} & \textbf{Scheduler} & \textbf{Max Length} & \textbf{Warmup Ratio} \\
    \midrule
    \multirow{6}{*}{Detector}
    & RoBERTa    & 6,000 & 6,000 & 5   & 32 & 5e-5  & linear & 512 & 0 \\
    & GPT2       & 6,000 & 6,000 & 5   & 32 & 5e-5  & linear & 512 & 0 \\
    & BERT       & 6,000 & 6,000 & 5   & 32 & 5e-5  & linear & 512 & 0 \\
    & MPU        & 6,000 & 6,000 & 5   & 32 & 5e-5  & linear & 512 & 0 \\
    & CheckGPT   & 6,000 & 6,000 & 5   & 32 & 2e-4  & cosine & 512 & 0 \\
    & GLTR       & 6,000 & 6,000 & -   & -  & -     & -      & 512 & - \\
    \hline
    Encoder-decoder & BERT-BERT   & 15,000 & 30,000 & 1.8 & 16 & 5e-5  & linear & 512 & 0.1 \\
    \hline
    Evader              & GradEscape  & 0     & 9,000  & 5   & 64 & 5e-5  & linear & 512 & 0 \\
    \bottomrule
\end{tabular}
}
\end{table*}

\begin{table}[!tbp]
\centering
\caption{Detector performance without attacks.
$\text{ER}_\text{b}$ represents the baseline evasion rate, which indicates how many machine-generated texts are identified as human-generated.}
\label{tab:detector_performance}
\resizebox{0.9\columnwidth}{!}{
\begin{tabular}{lcccccc}
\toprule
         & \multicolumn{2}{c}{RoBERTa} & \multicolumn{2}{c}{GPT2} & \multicolumn{2}{c}{BERT} \\
         \cmidrule(lr){2-3} \cmidrule(lr){4-5} \cmidrule(lr){6-7}
Dataset  & F1 & $\text{ER}_\text{b}$ & F1 & $\text{ER}_\text{b}$ & F1 & $\text{ER}_\text{b}$\\ 
\midrule
GROVER   & 0.921    & 0.016     & 0.919    & 0.012    & 0.906    & 0.001   \\
HC3      & 0.995    & 0.004     & 0.985    & 0.000    & 0.992    & 0.001   \\
GPA      & 0.994    & 0.000     & 0.992    & 0.000    & 0.996    & 0.002   \\
GPTWiki  & 0.983    & 0.000     & 0.972    & 0.002    & 0.987    & 0.003   \\
\bottomrule
\end{tabular}
}
\vspace{0.2cm}
\end{table}

\begin{compactitem}[$\bullet$]
    \item \textit{GROVER News~\cite{zellers2019defending}.}
    GROVER is a large CLM trained on RealNews~\cite{zellers2019defending} dataset using GPT3 as the base model.
    RealNews dataset is a large corpus (120GB) of news articles from Common Crawl~\cite{commoncrawl}.
    We contacted the authors to get the RealNews dataset.
    GROVER has not been trained using RLHF. Instead, it was trained by SFT using attributes including ``domain'', ``data'', ``authors'', and ``title'' as the prompt and news as the output.
    So that the user can generate news by inputting the above attributes.
    We use the largest version of GROVER, named GROVER-Mega, to create the machine-generated dataset.
    GROVER-Mega has 48 layers and 1.5 billion parameters.
    We sample RealNews to create the human-generated dataset.
    \item \textit{HC3~\cite{guo2023close}.}
    The Human ChatGPT Comparison Corpus (HC3) is a question-answering dataset collected to study the characteristics of ChatGPT's responses compared to human experts.
    It aims to understand how close ChatGPT's responses are to those of human experts across various domains, covering open-domain, financial, medical, legal, and psychological areas.
    HC3 dataset collects questions from these domains and their corresponding human answers, then uses ChatGPT (powered by GPT3.5) to generate the machine-generated dataset.
    \item \textit{GPA~\cite{liu2023check}.}
    The GPABenchmark is a cross-disciplinary corpus comprising human-written, GPT-written, GPT-completed, and GPT-polished research paper abstracts.
    It is introduced with the aim of training and evaluating detectors for identifying LLM-generated academic writing.
    The authors collect the human-generated dataset from arXiv and use ChatGPT (based on GPT3.5) to constrain the machine-generated dataset.
    In our paper, we only focus on GPT-written abstracts since they are more harmful than GPT-completed and GPT-polished abstracts.
    \item \textit{GPTWiki Intro~\cite{aaditya_bhat_2023}.}
    This dataset contains Wikipedia introductions and GPT-generated introductions for 150k topics.
    The authors use the title of the Wikipedia page and the first 7 words from the introduction paragraph as prompts.
    This dataset is generated by a GPT3 model fine-tuned for instruction-following.
\end{compactitem}

\mypara{Metrics}
We employ the following five metrics to estimate evaders' performance.
The evasion rate is used to measure attack utility.
Since the attacker hopes to constrain the syntactic and semantic similarity and ensure readability, we employ four extra text quality metrics: ROUGE, cos-sim, perplexity, and GRUEN.

\begin{compactitem}[$\bullet$]
    \item \textit{Evasion Rate.}
    The evasion rate refers to the fraction of machine-generated texts edited by evaders that are classified as human-written.
    More specifically given a machine-generated text dataset $\mathcal{T}_M$,
    \begin{equation}
        \text { Evasion Rate }=\frac{\sum_{i=0}^{\left|\mathcal{T}_M\right|} \mathbf{1}\left(D_V\left(\mathcal{F}_\theta\left(\mathcal{T}_{M, i}\right)\right) = y_{\text{human}}\right)}{\left|\mathcal{T}_M\right|},
    \end{equation}
    where $\mathbf{1}\{\mathcal{I}\}$ denotes the indicator function that is 1 when $\mathcal{I}$ is true and 0 when $\mathcal{I}$ is false.
    \item \textit{ROUGE~\cite{lin-2004-rouge}.}
    Since attack utility and text quality are in a trade-off, we need to synthesize the evasion rate with text quality metrics to evaluate evaders.
    ROUGE is a text quality metric measuring syntactic similarity between texts before and after being edited by evaders.
    It computes the fraction of text overlap.
    ROUGE has several variants.
    In this paper, we employ ROUGE-Lsum as the calculation method.
    \item \textit{Cos-sim.}
    This metric is used to compute semantic similarities.
    We first use Google's \textit{universal sentence encoder} (USE)~\cite{cer2018universal} to generate embeddings for texts before and after editing, and then compute their cosine distance.
    \item \textit{Perplexity.}
    Perplexity is a metric to measure text readability.
    A low perplexity indicates that the model is more confident in its predictions, and the text is more fluent and easier to read.
    Mathematically, the perplexity of a probabilistic model on a piece of text is defined as:
    \begin{equation}
        \text { PPL }=2^{-\frac{1}{N_w} \sum_{i=1}^{N_w} \log _2 P\left(w_i \mid w_1, w_2, \ldots, w_{i-1}\right)}.
    \end{equation}
    Here, $N_w$ is the number of words; $P\left(w_i \mid w_1, w_2, \ldots, w_{i-1}\right)$ is the conditional probability of word $w_i$ given the preceding word.
    \item \textit{GRUEN~\cite{zhu2020gruen}.}
    GRUEN is also used to measure text readability.
    Unlike perplexity, GRUEN normalizes its values between 0 and 1.
    GRUEN evaluates text quality from multiple perspectives, including ``grammaticality'', non-redundancy, discourse, focus, structure, and coherence.
\end{compactitem}

\begin{figure*}[!tbp]
    \centering
    \includegraphics[width=0.85\linewidth]{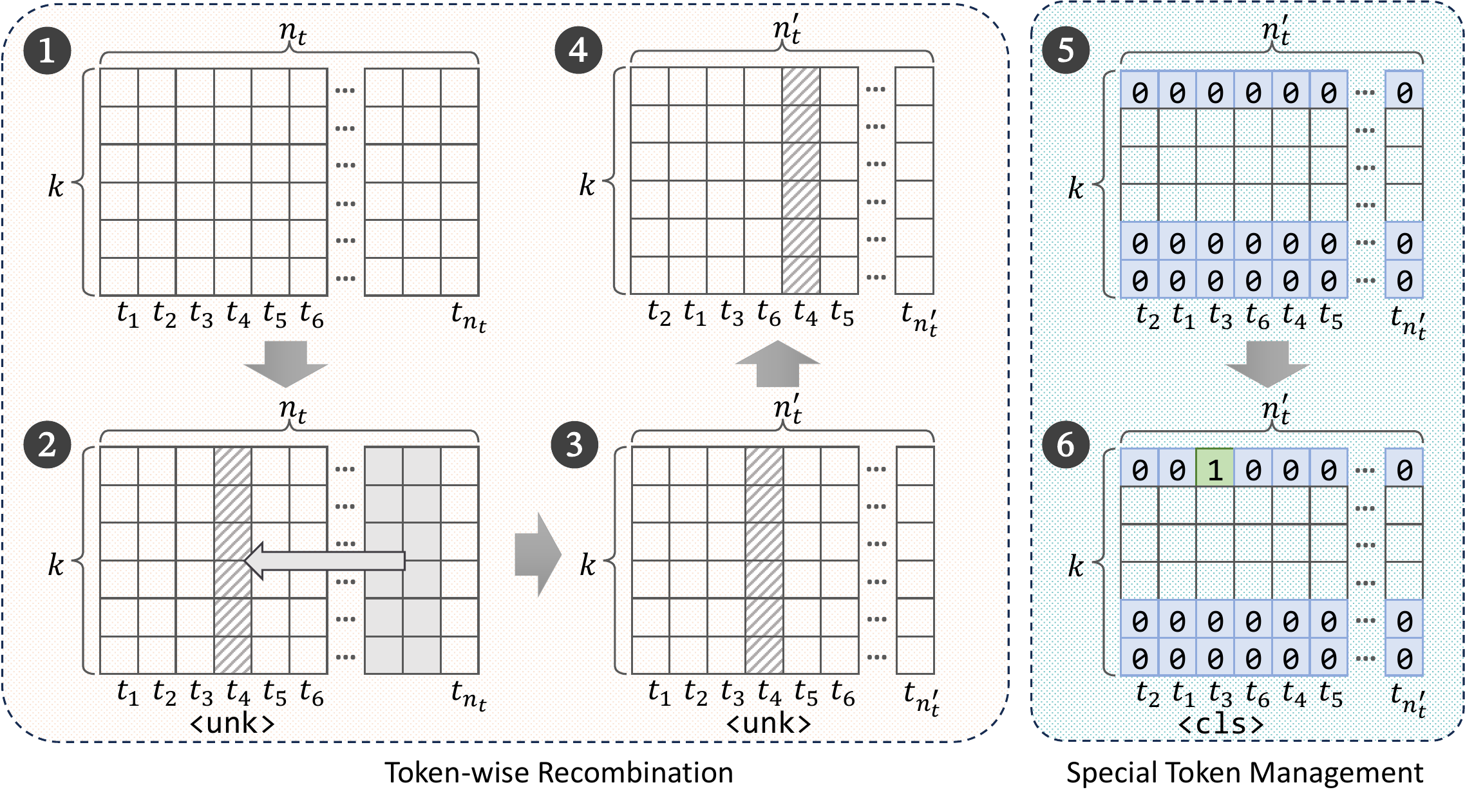}
    \caption{Detailed steps of probability matrix post-processing.}
    \label{fig:post-processing}
\end{figure*}

\mypara{Detectors Setup}
Our evaluation includes three basic detectors (RoBERTa, GPT2, and BERT), two advanced detectors (MPU and CheckGPT), and one statistic-based detector (GLTR).
The hyperparameters of detectors' training are detailed in~\autoref{tab:hyperparameters}.

\begin{compactitem}[$\bullet$]
\item \textit{Basic Detectors.}
We fine-tune these models on each dataset using 12,000 training samples and 4,000 test samples with an equal split between LLM-generated and human-generated texts.
The training uses a batch size of 32, 5 epochs, a learning rate of 5e-5, and early stopping to prevent overfitting.
\item \textit{MPU~\cite{tian2023multiscale}.}
This detector is designed to improve detection accuracy on short texts by leveraging multi-scale data augmentation and an additional PU loss. We follow the original paper's configurations for data augmentation and the PU loss.
For training arguments, we set the same as those of basic detectors.
\item \textit{CheckGPT~\cite{liu2023check}.}
This detector freezes the LM and adds a BiLSTM head to enhance training efficiency.
We set the learning rate to 2e-4 and use a cosine learning rate scheduler as described in the original paper.
Other training arguments are consistent with those for basic detectors.
\item \textit{GLTR~\cite{gehrmann2019gltr}.}
GLTR collects probability ranks for each word in a text to create word-rank features.
We use GPT2-XL to compute these probabilities and train a logistic regression classifier for classification.
We use grid search with an L2 penalty to find the optimal model.
\end{compactitem}

\mypara{Evaders Setup}
We select two perturbation-based evaders (RP and DFRFooler) and two paraphrase-based evaders (DIPPER and SentPara) as baselines.
\begin{compactitem}[$\bullet$]
\item \textit{RP~\cite{pu2023deepfake}.}
This method randomly replaces words with synonyms.
We vary the number of replacements in \{10, 15, 20, 25, 30, 35, 40\}.
\item \textit{DFTFooler~\cite{pu2023deepfake}.}
DFTFooler replaces words with the highest prediction probabilities.
We choose GPT2-XL to compute these probabilities.
The number of replacements considered is the same as RP.
\item \textit{DIPPER~\cite{krishna2024paraphrasing}.}
This evader paraphrases target texts in an end-to-end manner.
We use the fine-tuned T5-XXL model released by the authors.
We evaluate four LEX-ORD settings: \{L20-O0, L40-O0, L60-O0, L60-O60\}, which control lexicon and order consistency.
\item \textit{SentPara~\cite{sadasivan2023can}.}
SentPara uses Parrot~\cite{prithivida2021parrot} toolkit's small paraphrase models to paraphrase texts sentence-by-sentence, then concatenates the results.
We use the fine-tuned T5-Base paraphrase and evaluate all five settings controlling Levenshtein distance.
\end{compactitem}

\section{Probability Matrix Post-processing}
\label{app:post-processing}

In \method, the probability matrix $P$ output by the evader is fed into the text encoder and the detector model.
We refer to both the text encoder and the detector model collectively as \textit{downstream models}.
Even though the evader model and the downstream model utilize the same type of tokenizers, there may still be structural inconsistencies between $P$ and $W_{emb}$.
To bridge this gap, we employ a method named probability matrix post-processing.
This process comprises two phases, as illustrated in~\autoref{fig:post-processing}.
The first phrase is \textit{token-wise recombination}, where we handle excess token columns and rearrange the order of other token columns.
The second phase is \textit{special token management}, where we mask the evader's special tokens and add the special tokens required by the downstream model.

\mypara{Token-wise Recombination}
The number and order of tokens in the tokenizer of the evader model may differ from those in the downstream model's tokenizer.
To address discrepancies in token quantity, we add the values from the excess token columns in $P^{k\times n_{t}}$ to the \sptoken{<unk>} column and remove these excess token columns, as shown in step \X2.
By doing so, we transform the act of the evader model predicting excess tokens into an increased probability of predicting \sptoken{<unk>}.
To resolve token order inconsistency, we apply the token-remapping algorithm proposed by Bagdasaryan et al.~\cite{bagdasaryan2022spinning} to reorder the token columns.
Finally, in step \X4, we get a resized and recorded probability matrix $P^{k\times n_{t}'}$.

\mypara{Special Token Management}
In step \X5, we mask the rows in $P^{k\times n_{t}'}$ based on the evader's input token IDs.
The input format for BART is structured as: $x = {\text{\sptoken{<cls>}}, x_1, x_2, \dots, x_m, \text{\sptoken{<sep>}}, \text{\sptoken{<pad>}}}$.
The output also adheres to this format.
In this example, there are three special tokens in the sequence, which means $k = m + 3$.
We zero out the rows corresponding to special tokens to eliminate their influence in the downstream model's output.
In step \X6, we set the positions for the essential special tokens required by the downstream model to one.

\begin{table}[!tbp]
    \centering
    \caption{Open model attack results against MPU.}
    \label{tab:attack_results_mpu}
    \resizebox{\columnwidth}{!}{
    \begin{tabular}{llccccc}
    \toprule
    Dataset & Evader & ROUGE & cos-sim & PPL$\downarrow$ & GRUEN & ER\\
    \midrule
    \multirow{5}{*}{GROVER} & \method & 0.901 & 0.960 & 40.5 & 0.684 & \textbf{0.966} \\
     & RP & 0.915 & \textbf{0.986} & 50.2 & 0.584 & 0.789 \\
     & DFTFooler & 0.900 & 0.982 & 76.6 & 0.524 & 0.964 \\
     & DIPPER & 0.875 & 0.969 & \textbf{23.3} & \textbf{0.764} & 0.963 \\
     & SentPara & 0.909 & 0.977 & 40.8 & 0.661 & 0.248 \\
    \midrule
    \multirow{5}{*}{HC3} & \method & 0.923 & \textbf{0.986} & 19.5 & 0.664 & \textbf{0.766} \\
     & RP & 0.896 & 0.977 & 32.8 & 0.614 & 0.171 \\
     & DFTFooler & 0.895 & 0.975 & 46.4 & 0.589 & 0.230 \\
     & DIPPER & 0.901 & 0.965 & \textbf{11.1} & \textbf{0.799} & 0.005 \\
     & SentPara & 0.906 & 0.981 & 16.2 & 0.748 & 0.054 \\
    \midrule
    \multirow{5}{*}{GPA} & \method & 0.902 & 0.963 & 31.4 & 0.614 & \textbf{0.404} \\
     & RP & 0.908 & 0.976 & 37.6 & 0.662 & 0.161 \\
     & DFTFooler & 0.905 & \textbf{0.979} & 49.9 & 0.636 & 0.173 \\
     & DIPPER & 0.873 & 0.959 & \textbf{15.5} & \textbf{0.823} & 0.005 \\
     & SentPara & 0.922 & 0.979 & 20.7 & 0.779 & 0.012 \\
    \midrule
    \multirow{5}{*}{GPTwiki} & \method & 0.909 & 0.975 & 24.7 & 0.700 & \textbf{0.616} \\
     & RP & 0.892 & 0.983 & 35.7 & 0.603 & 0.467 \\
     & DFTFooler & 0.893 & 0.981 & 50.2 & 0.560 & 0.599 \\
     & DIPPER & 0.903 & 0.963 & \textbf{15.1} & \textbf{0.770} & 0.008 \\
     & SentPara & 0.906 & \textbf{0.983} & 30.2 & 0.658 & 0.279 \\
    \bottomrule
    \end{tabular}
    }
    \vspace{0.2cm}
\end{table}

\begin{table}[tbp]
    \centering
    \caption{Open model attack results against CheckGPT.}
    \label{tab:attack_results_checkgpt}
    \resizebox{\columnwidth}{!}{
    \begin{tabular}{llccccc}
    \toprule
    Dataset & Evader & ROUGE & cos-sim & PPL$\downarrow$ & GRUEN & ER\\
    \midrule
    \multirow{5}{*}{GROVER} & \method & 0.891 & 0.963 & 46.3 & 0.674 & \textbf{0.972} \\
     & RP & 0.915 & \textbf{0.986} & 50.2 & 0.584 & 0.061 \\
     & DFTFooler & 0.900 & 0.982 & 76.6 & 0.524 & 0.104 \\
     & DIPPER & 0.875 & 0.969 & \textbf{23.3} & \textbf{0.764} & 0.901 \\
     & SentPara & 0.909 & 0.977 & 40.8 & 0.661 & 0.018 \\
    \midrule
    \multirow{5}{*}{HC3} & \method & 0.903 & 0.977 & 21.2 & 0.687 & \textbf{0.704} \\
     & RP & 0.896 & 0.977 & 32.8 & 0.614 & 0.562 \\
     & DFTFooler & 0.895 & 0.975 & 46.4 & 0.589 & 0.545 \\
     & DIPPER & 0.901 & 0.965 & \textbf{11.1} & \textbf{0.799} & 0.038 \\
     & SentPara & 0.906 & \textbf{0.981} & 16.2 & 0.748 & 0.152 \\
    \midrule
    \multirow{5}{*}{GPA} & \method & 0.904 & 0.971 & 31.5 & 0.676 & \textbf{0.858} \\
     & RP & 0.908 & 0.976 & 37.6 & 0.662 & 0.503 \\
     & DFTFooler & 0.905 & \textbf{0.979} & 49.9 & 0.636 & 0.468 \\
     & DIPPER & 0.873 & 0.959 & \textbf{15.5} & \textbf{0.823} & 0.024 \\
     & SentPara & 0.922 & 0.979 & 20.7 & 0.779 & 0.093 \\
    \midrule
    \multirow{5}{*}{GPTwiki} & \method & 0.925 & 0.982 & 21.6 & 0.720 & \textbf{0.534} \\
     & RP & 0.892 & 0.983 & 35.7 & 0.603 & 0.529 \\
     & DFTFooler & 0.893 & 0.981 & 50.2 & 0.560 & 0.469 \\
     & DIPPER & 0.903 & 0.963 & \textbf{15.1} & \textbf{0.770} & 0.037 \\
     & SentPara & 0.906 & \textbf{0.983} & 30.2 & 0.658 & 0.412 \\
    \bottomrule
    \end{tabular}
    }
\end{table}

\section{Extra Open Model Attack Results}

\subsection{Evade Basic Detectors}
\label{app:extra_white_box}

\autoref{tab:detector_performance} shows detectors' performance before being attacked.
We can see that all detectors have low baseline evasion rates ($<2\%$).

\autoref{fig:white_box_gpa} and~\autoref{fig:white_box_gptwiki} show open model attack results on GPA dataset and GPTWiki dataset, respectively.
\autoref{tab:attack_examples} illustrates evasion examples of \method.
We observe that \method behaves differently across various datasets.
On the GROVER dataset, which is relatively easy to attack, \method altered only a few words, resembling a perturbation-based evader.
Conversely, on the HC3 dataset, which is more challenging to attack, \method reorganized the text structure, demonstrating characteristics of a paraphrase-based evader.

\subsection{Evade Advanced Detectors}
\label{sec:open_advanced_attack}

In the previous experiments, the victim detectors were vanilla fine-tuned LMs.
Recently, some researchers have proposed improved deep learning-based detectors for \aigt detection.
MPU~\cite{tian2023multiscale} alters the training process by incorporating PU loss, while CheckGPT~\cite{liu2023check} modifies model architecture by adding a BiLSTM head to the output side of the LM.
In this section, we show that our \method can handle both modifications on training loss and model architecture.

\mypara{Attack Setup}
We conduct attacks on MPU and CheckGPT across four datasets and compare \method results with our baseline evaders.
We adjust $\alpha$ and $\beta$ to ensure that \method's ROUGE score is around 0.9.
For baseline evaders, we choose the experimental results where the ROUGE score is closest to 0.9.

\mypara{Results}
The open attack results against MPU and CheckGPT are shown in~\autoref{tab:attack_results_mpu} and \autoref{tab:attack_results_checkgpt}, respectively.
We find that \method achieves higher evasion rates than all baselines against both MPU and CheckGPT.
Although DIPPER has the highest text quality among all evaders, it exhibits a low evasion rate.
Note that DIPPER is an 11B LLM; both its training and inference overhead are very high.
The text quality of \method is significantly better than that of perturbation-based evaders and also slightly better than SentPara, which is similar in size to \method.

\section{Extra Opaque Model Attack Results}

\begin{table}[tbp]
    \centering
    \caption{Opaque model attack results against MPU.}
    \label{tab:opaque_attack_results_mpu}
    \resizebox{\columnwidth}{!}{
    \begin{tabular}{llccccc}
    \toprule
    Dataset & Attack & ROUGE & cos-sim & PPL$\downarrow$ & GRUEN & ER\\
    \midrule
    \multirow{3}{*}{GROVER} & Query & 0.905 & 0.955 & \textbf{32.4} & \textbf{0.722} & 0.974 \\
     & Shadow & 0.917 & \textbf{0.981} & 33.9 & 0.720 & \textbf{0.996} \\
     & Open & 0.901 & 0.960 & 40.5 & 0.684 & 0.966 \\
    \midrule
    \multirow{3}{*}{HC3} & Query & 0.920 & 0.960 & 24.0 & \textbf{0.715} & 0.366 \\
     & Shadow & 0.902 & 0.967 & 21.4 & 0.695 & 0.300 \\
     & Open & 0.923 & \textbf{0.986} & \textbf{19.5} & 0.664 & \textbf{0.766} \\
    \midrule
    \multirow{3}{*}{GPA} & Query & 0.910 & \textbf{0.966} & 32.7 & \textbf{0.643} & 0.422 \\
     & Shadow & 0.886 & 0.927 & 37.0 & 0.590 & \textbf{0.430} \\
     & Open & 0.902 & 0.963 & \textbf{31.4} & 0.614 & 0.404 \\
    \midrule
    \multirow{3}{*}{GPTwiki} & Query & 0.921 & 0.975 & 29.6 & 0.668 & 0.568 \\
     & Shadow & 0.916 & \textbf{0.980} & 25.4 & 0.666 & 0.472 \\
     & Open & 0.909 & 0.975 & \textbf{24.7} & \textbf{0.700} & \textbf{0.616} \\
    \bottomrule
    \end{tabular}
    }
    \vspace{0.2cm}
\end{table}

\subsection{Evade Advanced Detectors}
\label{sec:eval_opaque_advanced_detectors}

In this section, we demonstrate the effectiveness of \method against advanced detectors in opaque model attack scenarios.
We utilize the same attack setup as described in~\autoref{sec:open_advanced_attack}.
We execute query and shadow dataset attacks against MPU and CheckGPT across four datasets.
For the query attacks, we choose 4000 samples as the query dataset.
The experimental outcomes for MPU and CheckGPT are presented in~\autoref{tab:opaque_attack_results_mpu} and~\autoref{tab:opaque_attack_results_checkgpt}, respectively.
We can find that opaque model attacks achieve comparable or even higher evasion rates than open model attacks with similar text quality.
The evasion rates of query and shadow dataset attacks against CheckGPT are higher than those of open model attacks, primarily because CheckGPT freezes most of the model parameters, updating only the parameters of the classification head, which helps prevent overfitting of the surrogate model.

\begin{table}[tbp]
    \centering
    \caption{Opaque model attack results against CheckGPT.}
    \label{tab:opaque_attack_results_checkgpt}
    \resizebox{\columnwidth}{!}{
    \begin{tabular}{llccccc}
    \toprule
    Dataset & Attack & ROUGE & cos-sim & PPL$\downarrow$ & GRUEN & ER\\
    \midrule
    \multirow{3}{*}{GROVER} & Query & 0.897 & 0.950 & 54.4 & 0.654 & \textbf{1.000} \\
     & Shadow & 0.892 & 0.956 & \textbf{43.6} & \textbf{0.684} & \textbf{1.000} \\
     & Open & 0.891 & \textbf{0.963} & 46.3 & 0.674 & 0.972 \\
    \midrule
    \multirow{3}{*}{HC3} & Query & 0.891 & 0.972 & 18.8 & \textbf{0.688} & \textbf{1.000} \\
     & Shadow & 0.900 & \textbf{0.983} & \textbf{18.4} & 0.649 & \textbf{1.000} \\
     & Open & 0.903 & 0.977 & 21.2 & 0.687 & 0.704 \\
    \midrule
    \multirow{3}{*}{GPA} & Query & 0.915 & 0.946 & 34.4 & \textbf{0.727} & \textbf{1.000} \\
     & Shadow & 0.897 & 0.952 & 36.1 & 0.683 & \textbf{1.000} \\
     & Open & 0.904 & \textbf{0.971} & \textbf{31.5} & 0.676 & 0.858 \\
    \midrule
    \multirow{3}{*}{GPTwiki} & Query & 0.905 & 0.960 & 27.0 & 0.648 & \textbf{1.000} \\
     & Shadow & 0.896 & 0.945 & 30.0 & 0.619 & \textbf{1.000} \\
     & Open & 0.925 & \textbf{0.982} & \textbf{21.6} & \textbf{0.720} & 0.534 \\
    \bottomrule
    \end{tabular}
    }
\end{table}

\subsection{Evade Statistic-based Detectors}
\label{sec:attack_statistic}

Apart from deep learning-based detectors, statistic-based detectors are also used for post-hoc detection.
Since \method updates the evader based on the detector's gradients, a concern with \method is whether it can evade statistic-based detectors.
In this section, we demonstrate that \method can also tackle statistic-based detectors by constructing a deep learning-based surrogate model.

\begin{table}[!tbp]
    \centering
    \caption{\method's query attack results against GLTR.}
    \label{tab:gltr_attack}
    \resizebox{\columnwidth}{!}{
    \begin{tabular}{lcccccc}
    \toprule
    Dataset & ROUGE & cos-sim & PPL$\downarrow$ & GRUEN & ER$_\text{b}$ & ER \\
    \midrule
    GROVER & 0.911 & 0.973 & 32.5 & 0.721 & 0.263 & 0.558 \\
    HC3 & 0.890& 0.979 & 19.3 & 0.667 & 0.047 & 0.724 \\
    GPA & 0.934 & 0.973 & 36.0 & 0.613 & 0.132 & 0.656 \\
    GPTWiki & 0.923 & 0.959 & 34.0 & 0.608 & 0.031 & 0.526 \\
    \bottomrule
    \end{tabular}
    }
    \vspace{0.2cm}
\end{table}

\mypara{Attack Setup}
We select GLTR~\cite{gehrmann2019gltr} as the victim detector since it performs the best among existing statistic-based detectors~\cite{he2023mgtbench}.
GLTR leverages the insight that decoding strategies tend to favor tokens assigned high probabilities by the LM.
Utilizing a backend LM, GLTR extracts features based on the number of tokens within the Top-10, Top-100, and Top-1000 ranks as determined by the token probability distributions.
We use GPT2-XL as the backend LM.
We choose 2000 samples to form the query dataset, and then fine-tune a GPT2-Base model as the surrogate model using the results of these queries.
As in previous experiments, we adjust $\alpha$ and $\beta$ to target a ROUGE score around 0.9.

\mypara{Results}
The results, as presented in~\autoref{tab:gltr_attack}, indicate that \method boosts the average evasion rate by 0.5 compared to the pre-attack baseline.
Notably, even for datasets like HC3 and GPTWiki, where GLTR demonstrates strong performance, \method achieves evasion rates exceeding 50\%.
This suggests that DNN models are capable of imitating the underlying patterns of statistic-based detectors through querying.

\subsection{Evade Heterogeneous Detectors}
\label{app:evade_heterogeneous}
\begin{figure*}[!tbp]
    \centering
    \includegraphics[width=\textwidth]{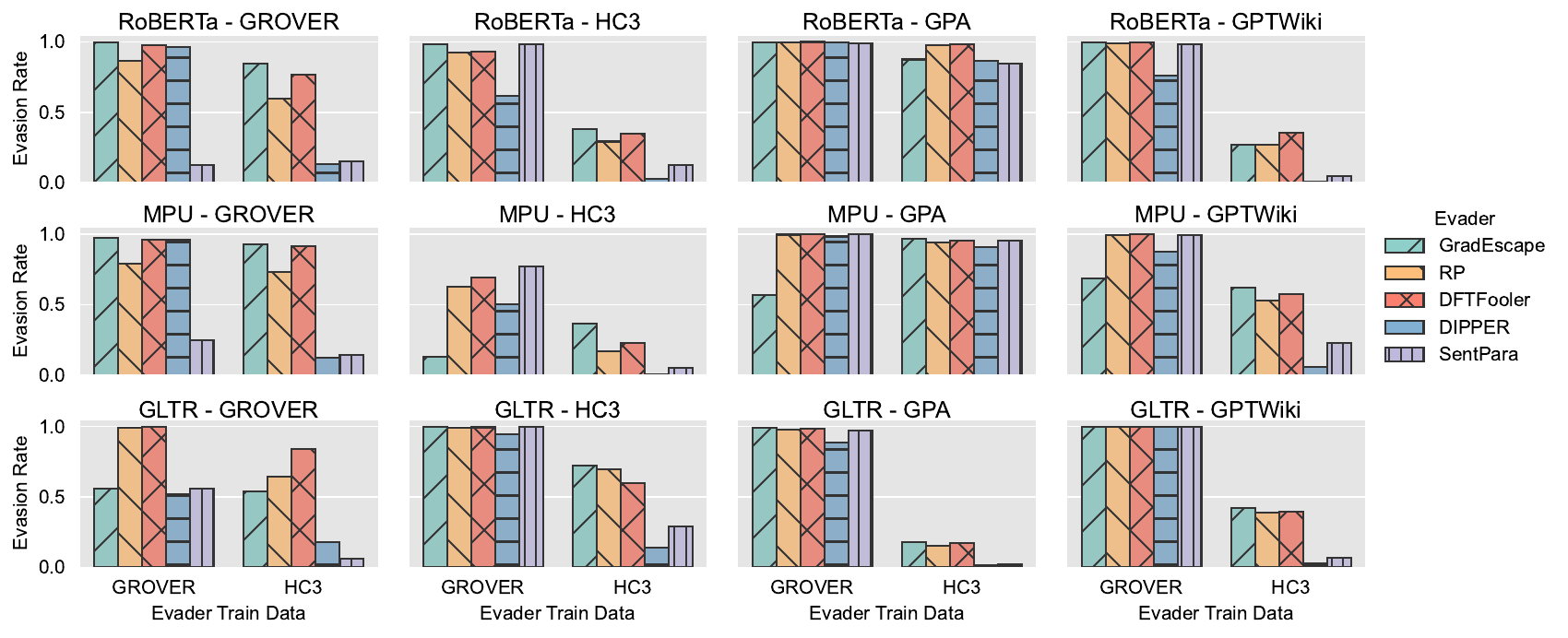}
    \caption{Transferability comparison with baselines under heterogeneous detector training. Each title ``DETECTOR - DATASET'' indicates that the victim model is a DETECTOR trained on the DATASET.}
    \label{fig:transfer_compare}
\end{figure*}

We compare \method with four baselines under scenarios where the training datasets of the evader and the detector vary.
Evasion rates are measured against three distinct detectors: RobERTa, MPU, and GLTR.
\autoref{fig:transfer_compare} presents the experimental results.
Out of 24 comparison groups, \method achieves the highest evasion rate in 16 cases.
Notably, when attacking the more challenging HC3 dataset \method ranks first in 9 out of 12 comparisons.
RP and DFTFooler are adaptive attacks for GLTR, and thus achieve high evasions against GLTR.
Even so, \method outperforms them when attacking GLTR in many cases.

\begin{table}[!tbp]
    \centering
    \caption{Performance of real-world detectors without attacks. Datasets in\hlcolor{lightgray}{gray}background are selected to attack.}
    \label{tab:test_real_world}
    \resizebox{\columnwidth}{!}{
    \begin{tabular}{ll|cccc}
    \toprule
    \textbf{Detector} & \textbf{Dataset} & \textbf{Precision} & \textbf{Recall} & \textbf{F1} & \textbf{Accuracy} \\
    \midrule

    & GROVER
    & 0.85
    & 0.28
    & 0.43
    & 0.62 \\

    \rowcolor{lightgray} \cellcolor{white}
    & HC3
    & 0.84
    & 0.99
    & 0.91
    & 0.91 \\

    & GPA
    & 0.61
    & 1.00
    & 0.76
    & 0.70 \\

    \rowcolor{lightgray} \cellcolor{white}
    \multirow{-4}{*}{Sapling}
    & GPTWiki
    & 0.72
    & 0.94
    & 0.82
    & 0.78 \\
    \midrule
    
    & GROVER
    & 1.00
    & 0.14
    & 0.25
    & 0.64 \\

    \rowcolor{lightgray} \cellcolor{white}
    & HC3 
    & 1.00
    & 0.97
    & 0.99
    & 0.98 \\

    \rowcolor{lightgray} \cellcolor{white}
    & GPA
    & 1.00
    & 0.99
    & 0.99
    & 0.99 \\

    \multirow{-4}{*}{Scribbr} 
    & GPTWiki
    & 0.93
    & 0.14
    & 0.25
    & 0.57 \\
    \bottomrule
    \end{tabular}
    }
    \vspace{0.2cm}
\end{table}

\section{Extra Real-world Experimental Results}
\label{app:extra_real_world}

The detection performance of Sapling and Scribbr without attacks is shown in~\autoref{tab:test_real_world}.
We choose two datasets for each real-world detector to conduct our \method attack.
\autoref{fig:tokenizer_inference_sapling} illustrates tokenizer inference attack observations against Sapling.
\autoref{fig:sapling_attack_example} and~\autoref{fig:scribbr_attack_example} show attack examples on HC3 dataset against Sapling and Scribbr, respectively.

\autoref{tab:real_world_comparison} presents the comparison with our baselines.
For each baseline, we adjust the knob so that the ROUGE scores are similar to those of the texts modified by \method.
As shown, \method achieves state-of-the-art evasion rates on both Sapling and Scribbr.
Although SentPara shows a greater $\overline{\text{DC}}$ decrease when attacking Scribbr with GPA dataset, \method's DC values are concentrated around 0.4, while SentPara's DC values are concentrated around 0.1.
\method affects more examples and attains a higher evasion rate.

\begin{table*}[!tbp]
    \centering
    \caption{Real-world comparison with baselines.}
    \label{tab:real_world_comparison}
    \resizebox{0.9\textwidth}{!}{
    \begin{tabular}{ll|ccccc|ccccc}
    \toprule
    & & \multicolumn{5}{c}{Evasion Rate} & \multicolumn{5}{c}{$\overline{\text{DC}}\downarrow$} \\
    \cmidrule(lr){3-7} \cmidrule(lr){8-12}
    Detector  & Train Data & RP    & DFTFooler & DIPPER & SentPara & \method & RP     & DFTFooler & DIPPER & SentPara & \method \\
    \midrule
    \multirow{2}{*}{Sapling} 
              & HC3       & 0.080  & 0.121     & 0.397  & 0.251 & \textbf{0.448}  & 0.913 & 0.878    & 0.609  & 0.745  & \textbf{0.550}  \\
              & GPTWiki   & 0.392 & 0.462     & 0.332  & 0.266  & \textbf{0.578}  & 0.602  & 0.535     & 0.669  & 0.729  & \textbf{0.422}  \\
    \midrule
    \multirow{2}{*}{Scribbr}
              & HC3       & 0.669 & 0.419     & 0.495  & 0.616 & \textbf{0.785}   & 0.467  & 0.622     & 0.559  & 0.481 & \textbf{0.378}   \\
              & GPA       & 0.520 & 0.354     & 0.520  & 0.564  & \textbf{0.658}  & 0.548  & 0.681     & 0.506  & \textbf{0.471}  & 0.496  \\
    \bottomrule
  \end{tabular}
  }
\end{table*}

\begin{table}[!tbp]
    \centering
    \caption{Percentage of GPT4-annotated major semantic changes. Lower is better.}
    \label{tab:gpt_annotation}
    \resizebox{0.9\columnwidth}{!}{
    \begin{tabular}{lcccc}
    \toprule
    \textbf{Evader} & \textbf{GROVER} & \textbf{HC3} & \textbf{GPA} & \textbf{GPTWiki} \\
    \midrule
    RP         & 13\% & 30\% & 31\% & 41\% \\
    DFTFooler  & 17\% & 47\% & 52\% & 44\% \\
    DIPPER     & 14\% & 11\% & 12\% & 12\% \\
    SentPara   & 25\% & 16\% & 12\% & 13\% \\
    GradEscape &  7\% & 29\% & 33\% & 21\% \\
    \bottomrule
    \end{tabular}
    }
\end{table}

\begin{figure}[!tbp]
\centering
\begin{mybox}{\textbf{\small{GPT4 Annotation Prompt}}}
\small{
Here are two texts. The first one is the original text, and the second one is the modified text.\\

Original text: \{\texttt{ORIGINAL\_TEXT}\}\\

Modified text: \{\texttt{MODIFIED\_TEXT}\}\\

Please tell me whether the modification changes the meaning of the original text or would confuse a reader. Please answer with "major," "minor," or "no change." "Major" means the modification causes key information to be missing, which could confuse readers; "minor" means the expression may be changed but can be correctly understood; and "no change" means the modification has little impact on the semantics.
}
\end{mybox}
\caption{The GPT4 annotation prompt for evaluating semantic consistency.}
\label{fig:gpt_annotation_prompt}
\end{figure}

\section{Semantic Consistency}
\label{app:semantic_consistency}

We assess the semantic consistency between modified and original texts by using both GPT-4-based annotation and human evaluation methods.

\mypara{GPT4 Annotation}
In particular, we employ \texttt{gpt-4o-2024-08-06} to determine whether the modifications made by evaders result in major semantic changes.
The annotation prompt is shown in~\autoref{fig:gpt_annotation_prompt}, and the percentage of major changes is reported in~\autoref{tab:gpt_annotation}.
For each evader on each dataset, we randomly sample 300 text pairs for annotation.
\method achieves semantic consistency levels between those of perturbation-based and paraphrase-based methods.
DIPPER, which utilizes an LLM for end-to-end paraphrasing, maintains high semantic consistency.
In contrast, perturbation-based methods are assessed as causing more major changes, especially on shorter datasets such as HC3 and GPTWiki.

\begin{figure*}[!tbp]
    \centering
    \includegraphics[width=\textwidth]{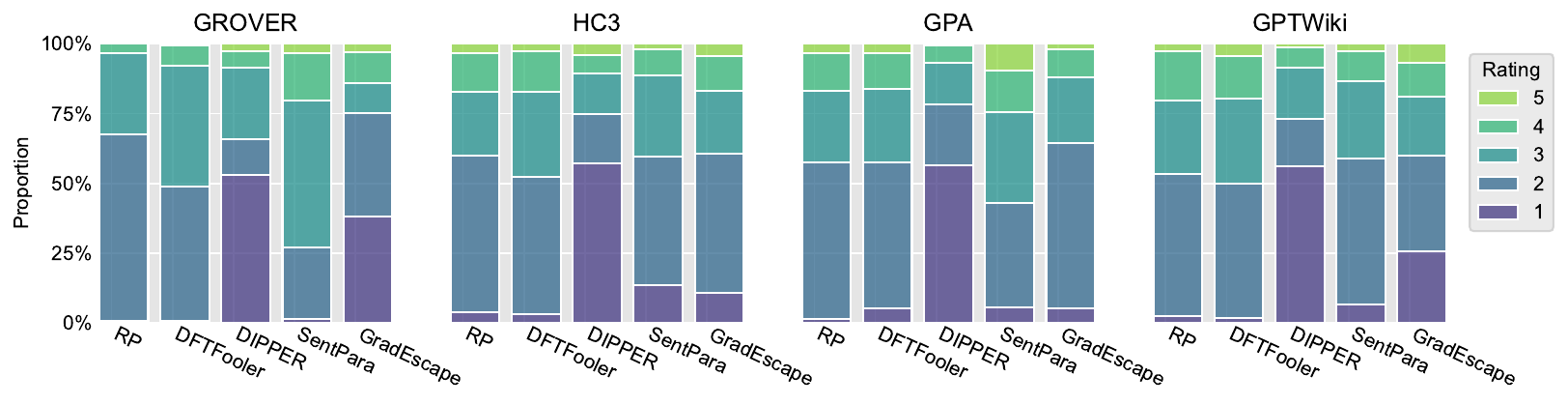}
    \caption{Human ratings of semantic changes. Higher scores indicate greater semantic distortion.}
    \label{fig:human_ratings}
\end{figure*}

\mypara{Human Evaluation}
We also develop a dedicated text semantic consistency rating platform (illustrated in~\autoref{fig:streamlit}) to facilitate human evaluation.
We hired 7 crowdworkers to rate semantic changes using this website.
Annotators are presented with pairs of original and modified texts and are asked to assign a score from 1 to 5, where a higher score indicates a greater semantic change.
We set a mandatory reading time of 25 seconds per rating, and annotators can choose to skip uncertain pairs.
For each evader on each dataset, we randomly choose 150 text pairs for human evaluation.
We pay each annotation USD 0.21 (XE rate as of 2025/05/19), with an estimated average hourly wage of USD 8.4 (40 pairs per hour), exceeding the minimum hourly wage in Hangzhou, China (USD 3.34, XE rate as of 2025/05/19).
This project was thoroughly reviewed and approved by the Science and Technology Ethics Committee of Zhejiang University, serving as our IRB.
To ensure rating consistency, we provided annotators with sample text pairs and reference scores from 1 to 5 annotated by authors. The experimental results are shown in~\autoref{fig:human_ratings}.
Human evaluation further confirms that \method's semantic consistency lies between perturbation-based and paraphrase-based methods.
SentPara performs worse in human evaluation, possibly because paraphrasing individual sentences may cause the combined text to become incoherent.
Interestingly, human annotators are generally more tolerant of perturbation-based evaders, suggesting that humans are less sensitive to changes involving the replacement of individual words.

\begin{figure*}[!tbp]
    \centering
    \includegraphics[width=\textwidth]{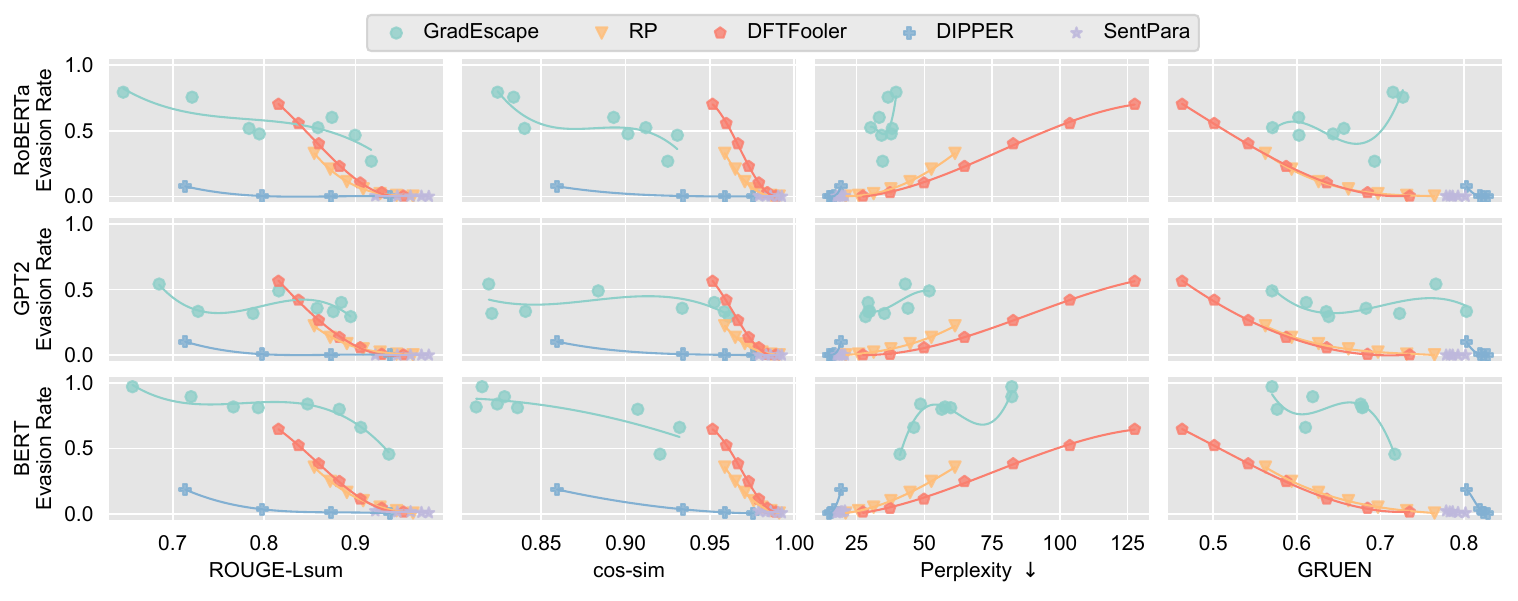}
    \caption{Evasion rates on GPA dataset.}
    \label{fig:white_box_gpa}
\end{figure*}

\begin{figure*}[!tbp]
    \centering
    \includegraphics[width=\textwidth]{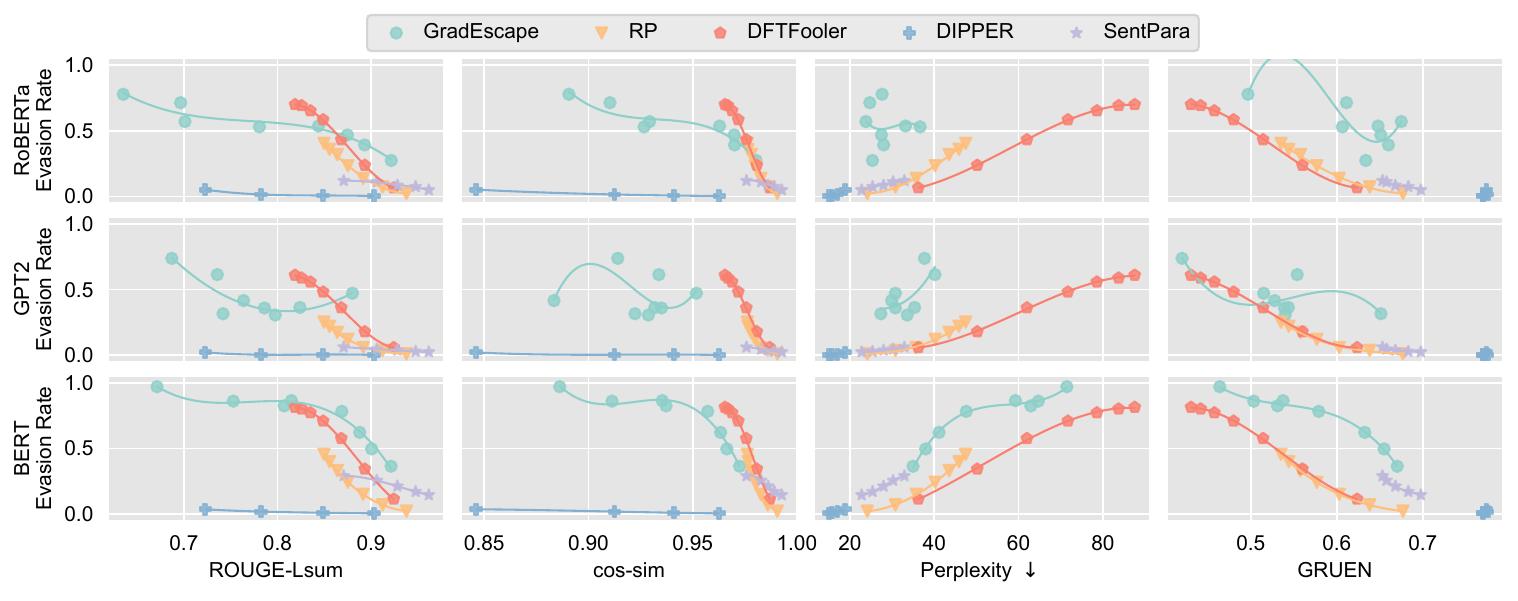}
    \caption{Evasion rates on GPTWiki dataset.}
    \label{fig:white_box_gptwiki}
\end{figure*}

\begin{table*}[!tbp]
    \centering
    \caption{Example outputs produced by \method. The differences are highlighted in \hl{red}.
    DC = Detection Confidence, which indicates the probability of being AI-generated predicted by the detector.}
    \label{tab:attack_examples}
    \resizebox{\textwidth}{!}{
    \begin{tabular}{lp{10cm}p{10cm}}
    \toprule
    \textbf{Dataset} & \textbf{Input} & \textbf{Output} \\
    \midrule
    GROVER
    &
    Print Article TOWN OF BELOIT - Two Beloit Memorial swimming teams dominated the 52nd annual Rock County meet Saturday at Ravenswood Pool, winning the boys' division and the girls' division. In the girls' competition, Beloit Memorial won all five events. The Cadets were led by senior Meredith Schultz, who won the 100 free and 100 breaststroke events. Marissa Cook (200 free) and Madison Wilson (200 individual medley and 100 backstroke) also were winners for \hl{Beloit} Memorial. Emma Sharkey finished first in the 200 free and won the 500 free. The Beloit boys finished with 596 points to win the team title. Logan Borowicz was a double winner for the Cadets. He won the 200 individual \hl{medley} and 100 \hl{fly}. John Ure won the 50 free and 100 free. Michael Aigner finished first in the 100 \hl{free}. Boys swimming Rock County meet - Beloit Memorial 596, Arrowhead 336, Rockford Christian 294, Whitewater 202, Independence 137, Hartford 107, Calumet 50, Beloit West 36 Boys relay - 1. Beloit Christian, 3:47.08; 2. Arrowhead, 3:47.28; 3. Rockford Christian, \hl{3:50.73;} 4. Rockford Christian, \hl{3}:51.28
    &
    Print Article TOWN OF BELOIT - Two Beloit Memorial swimming teams dominated the 52nd annual Rock County meet Saturday at Ravenswood Pool, winning the boys' division and the girls' division. In the girls competition, Beloit\colorbox{mycolor}{®} Memorial won all five events. The Cadets were led by senior Meredith Schultz, who won the 100 free and 100 breaststroke events. Marissa Cook (200 free) and Madison Wilson (200 individual medley and 100 backstroke) also were winners for \hl{Beloin} Memorial. Emma Sharkey finished first in the 200 free and won the 500 free. The Beloit boys finished with 596 points to win the team title. Logan Borowicz was a double winner for the Cadets. He won the 200 individual \hl{medLEY} and 100 \hl{free}. John Ure won the 50 free and 100 free. Michael Aigner finished first IN the 100 \hl{return}. Boys swimming Rock County meet - Beloit memorial 596, Arrowhead 336, Rockford Christian 294, Whitewater 202, Independence 137, Hartford 107, Calumet 50, Beloin West 36 Boys relay - 1. Beloit Christian, 3:47.08; 2. Arrowhead, 3.13.73; 3. Rockford Central, \hl{3::50.73?} 4. Rockf Christian, \hl{4}:51.28
    \\
    \addlinespace
    &
    \multicolumn{1}{r}{\parbox[t]{10cm}{\raggedleft \textit{DC: 0.842, PPL: 32.1, GRUEN: 0.475}}}
    &
    \multicolumn{1}{r}{\parbox[t]{10cm}{\raggedleft \textit{ROUGE: 0.948, cos-sim: 0.996, DC: 0.003, PPL: 56.2, GRUEN: 0.599}}}
    \\
    \midrule
    HC3
    &
    When you crack your knuckles or other joints, you are essentially releasing gas bubbles \hl{that have built up in the fluid that surrounds your joints. This can create a popping sound.} Cracking your knuckles on a regular basis is not likely to cause any long-term complications \hl{or damage to your joints.} However, \hl{it is possible to overdo it and cause temporary discomfort or swelling. So, it's probably a good idea to limit the amount of knuckle cracking you do}.
    &
    When you crack your knuckles or other joints, you are essentially releasing gas bubbles. \hl{Apparently, this can create popping sound.} Cracking your knuckle cracking on a regular basis is not likely to cause long-term complications. However, \hl{damage to your joints is possible to cause temporary discomfort. Thankfully, this effect is probably a good idea to limit the amount of knuckles cracking.}
    \\
    \addlinespace
    &
    \multicolumn{1}{r}{\parbox[t]{10cm}{\raggedleft \textit{DC: 0.999, PPL: 13.8, GRUEN: 0.865}}}
    &
    \multicolumn{1}{r}{\parbox[t]{10cm}{\raggedleft \textit{ROUGE: 0.741, cos-sim: 0.967, DC: 0.009, PPL: 31.8, GRUEN: 0.773}}}
    \\
    \midrule
    GPA
    &
    \hl{This paper} examines the politics of adversarial machine learning, which is the use of machine learning techniques to attack and defend against other machine learning algorithms. We argue that the political implications \hl{of adversarial machine learning are important} and underappreciated. Adversarial attacks can be used to undermine the performance of machine learning algorithms, potentially leading to harmful or biased decisions. Adversarial defenses, in turn, \hl{can be used} to protect against such attacks, but may also reinforce existing biases or inequalities. \hl{We} explore the ways in which adversarial machine learning interacts with broader political and social dynamics, including the distribution of power, the framing of security and risk, and the regulation of emerging technologies. We conclude by discussing the implications of our analysis \hl{for future research and policy initiatives}.
    &
    \hl{This is a paper} examines the politics of adversarial machine learning, which is the use of machine learning techniques to attack and defend against other machine learning algorithms. \hl{To this paper,} we argue that the political implications and underappreciated. Adversarial attacks can be used to undermine the performance of machinelearning algorithms, potentially leading to harmful or biased decisions. \hl{As a defense,} in turn, to protect against such attacks, but may also reinforce existing biases or inequalities. \hl{As part of our analysis for future research and policy initiatives.} \hl{To} explore the ways in which adversarial Machine learning interacts with broader political and social dynamics, including the distribution of power, the framing of security and risk, and the regulation of emerging technologies. \hl{To that end,} we conclude by discussing the implications of our analyses.
    \\
    \addlinespace
    &
    \multicolumn{1}{r}{\parbox[t]{10cm}{\raggedleft \textit{DC: 1.000, PPL: 12.5, GRUEN: 0.886}}}
    &
    \multicolumn{1}{r}{\parbox[t]{10cm}{\raggedleft \textit{ROUGE: 0.892, cos-sim: 0.958, DC: 0.390, PPL: 24.9, GRUEN: 0.795}}}
    \\
    \midrule
    GPTWiki
    &
    Radu Rosetti (Francized Rodolphe Rosetti; September 14, 1881 - November 25, 1957) was a French painter and printmaker\hl{. He} was associated with the Fauves and Cubists movements\hl{. Born in Bordeaux, France, to a family of wine merchants}, Rosetti studied at the Ecole des Beaux-Arts in Paris\hl{. In 1905 he exhibited his first painting at the Salon d'Automne.} The following year, he joined the Section d'Or group, whose members were associated with \hl{the Fauves} movement. In 1907 he exhibited with the group at the Salon des Independants\hl{. In 1908, Rosetti} exhibited at the Salon des Realites Nouvelles\hl{. He} joined the Societe Nationale de Peinture et de Sculpture in 1909\hl{, and became a member of} the Academie des Beaux-Arts in 1917. He lived in Montmartre from 1933 until his death.
    &
    Radu Rosetti (Francized Rodolphe Rosetti; September 14, 1881 - November 25, 1957) was a French painter and printmaker\hl{, and he} was associated with the Fauves and Cubists movements\hl{, and of course}, Rosetti studied at the Ecole des Beaux-Arts in Paris\hl{, and became a member of the Salon d'Automne at 1905.} The following year, he joined the Section d'Or group, whose members were associated with \hl{both the fauves} movement. In 1907 he exhibited with the group at the Salon des Independants\hl{, and in 1908 he} exhibited at Salon des Realites Nouvelles\hl{, and} joined the Societe Nationale de Peinture et de Sculpture in 1909\hl{. He joined} the Academie Des Beaux Arts in 1917. He lived in Montmartre from 1933 until his death.
    \\
    \addlinespace
    &
    \multicolumn{1}{r}{\parbox[t]{10cm}{\raggedleft \textit{DC: 0.999, PPL: 14.6, GRUEN: 0.761}}}
    &
    \multicolumn{1}{r}{\parbox[t]{10cm}{\raggedleft \textit{ROUGE: 0.901, cos-sim: 0.982, DC: 0.266, PPL: 22.8, GRUEN: 0.749}}}
    \\
    \bottomrule
    \end{tabular}
    }
\end{table*}

\begin{figure*}[!tbp]
    \centering
    \begin{subfigure}{.49\textwidth}
        \centering
        \includegraphics[width=\linewidth]{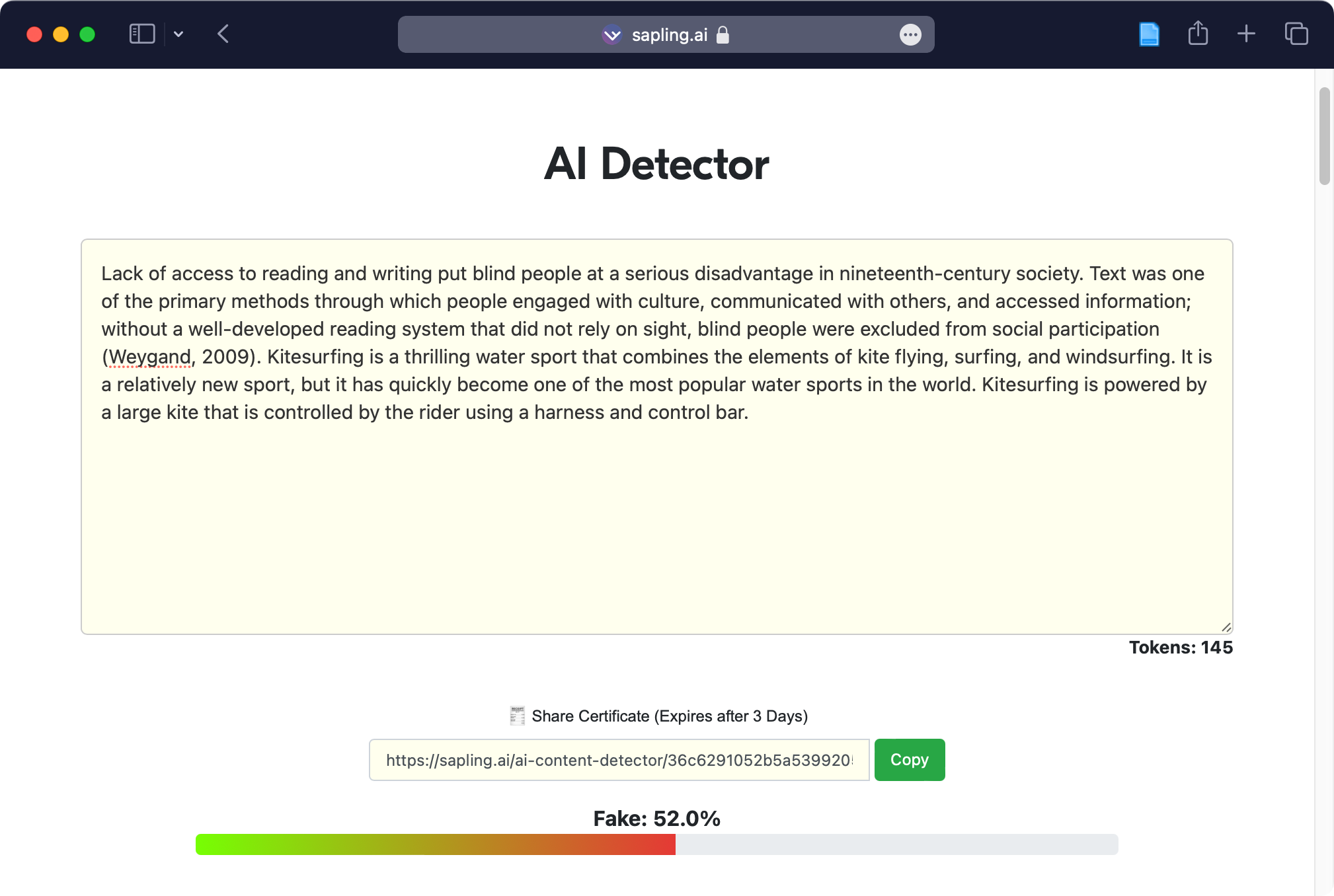}
        \caption{Original text.}
        \label{fig:sub1}
    \end{subfigure}%
    \hfill
    \begin{subfigure}{.49\textwidth}
        \centering
        \includegraphics[width=\linewidth]{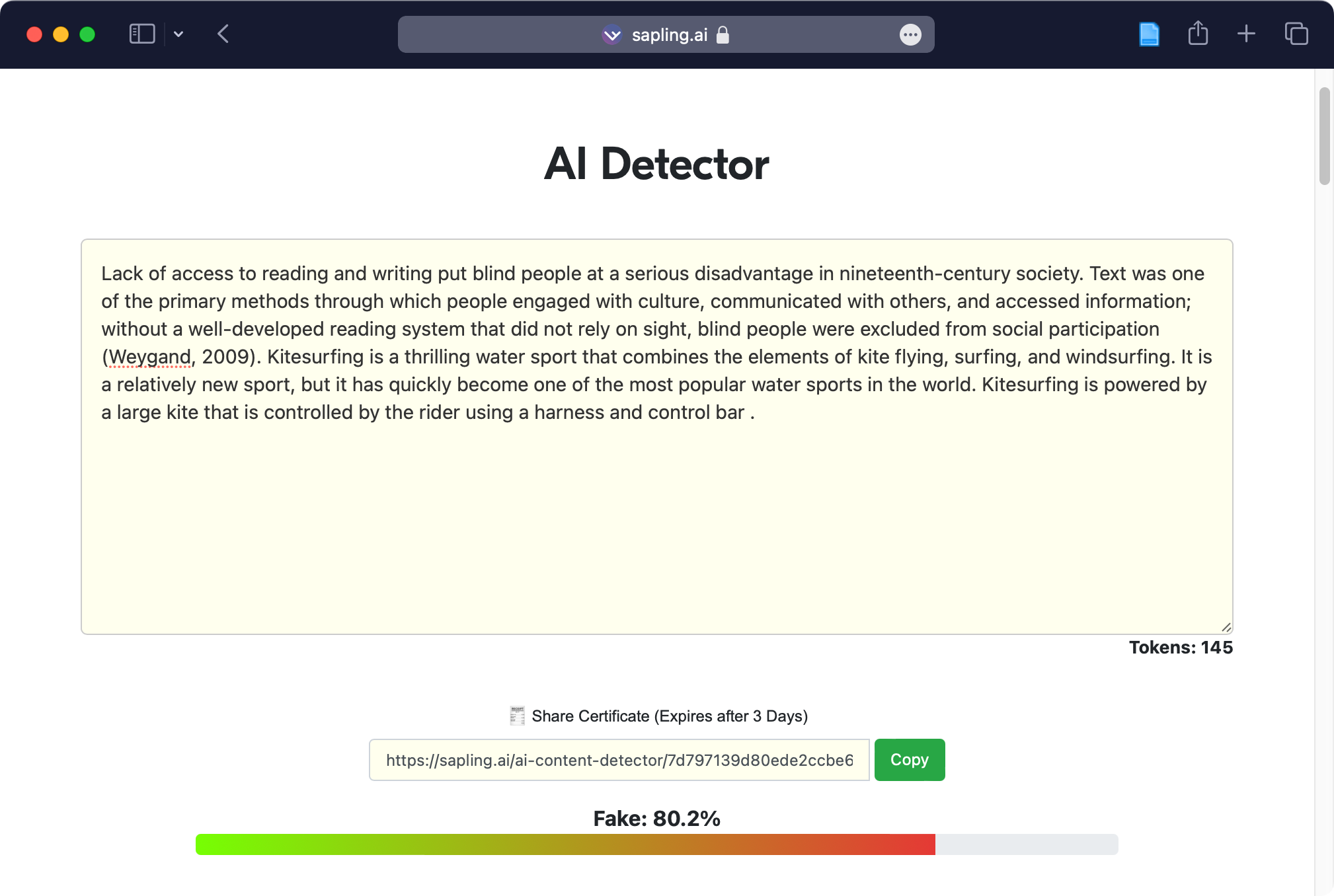}
        \caption{Inserting a space before the last period.}
        \label{fig:sub2}
    \end{subfigure}
    \begin{subfigure}{.49\textwidth}
        \centering
        \includegraphics[width=\linewidth]{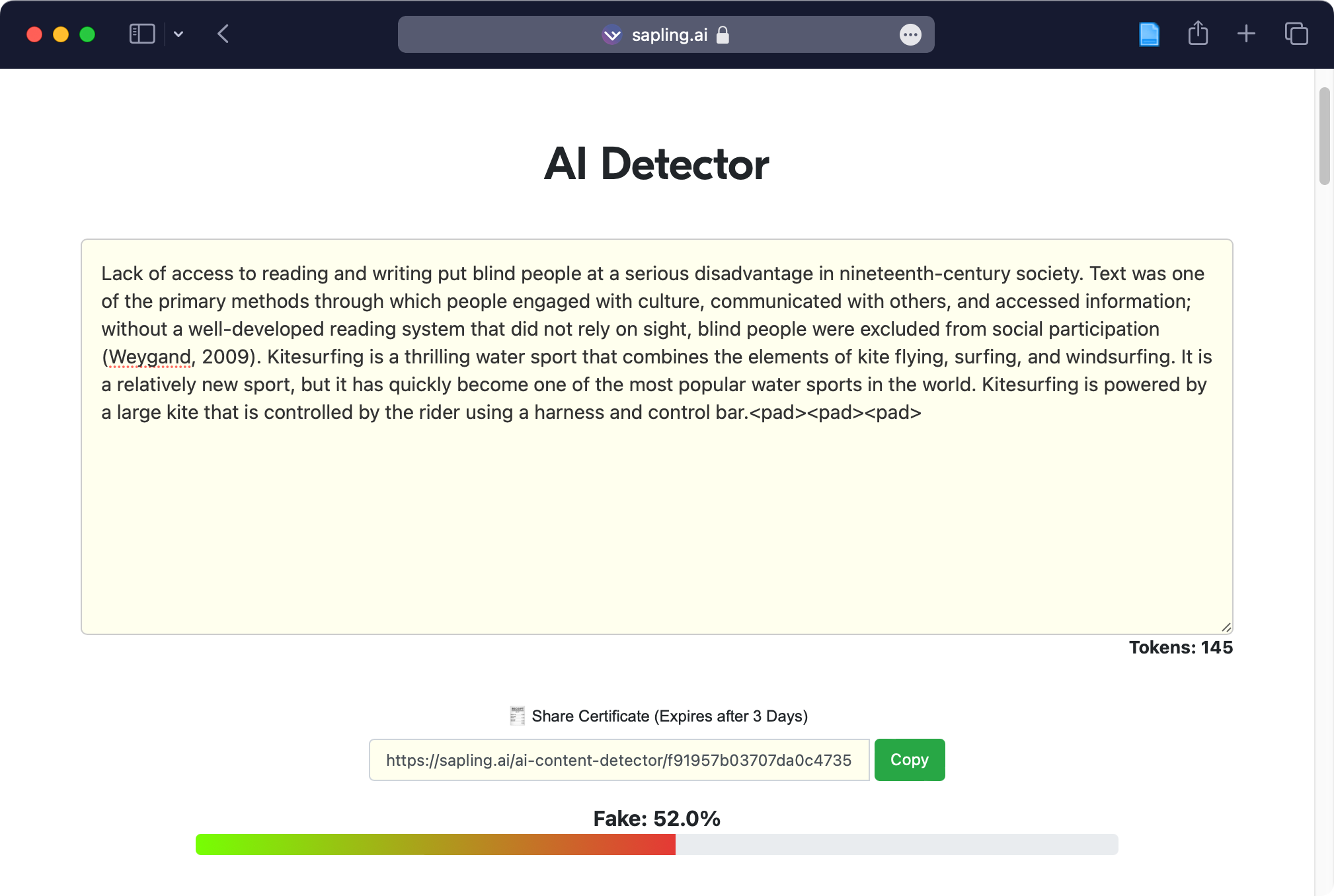}
        \caption{Appending \sptoken{<pad>}s to the text.}
        \label{fig:sub3}
    \end{subfigure}
    \hfill
    \begin{subfigure}{.49\textwidth}
        \centering
        \includegraphics[width=\linewidth]{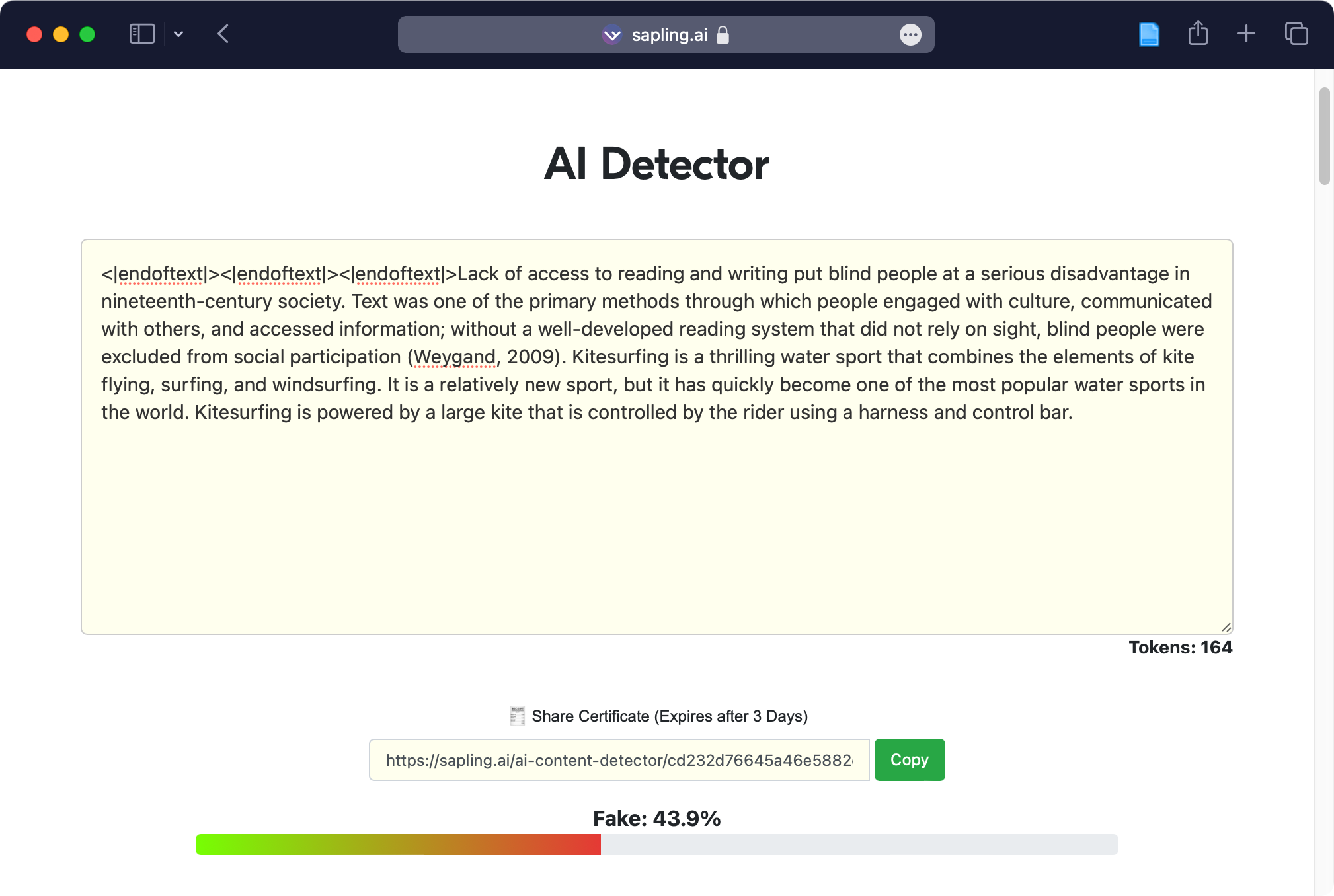}
        \caption{Prepending \sptoken{<|endoftext|>}s to the text.}
        \label{fig:sub4}
    \end{subfigure}
    \caption{Tokenizer inference attack against Sapling detector.}
    \label{fig:tokenizer_inference_sapling}
    \label{fig:test}
\end{figure*}

\begin{figure*}[!tbp]
    \centering
    \begin{subfigure}{.49\textwidth}
        \centering
        \includegraphics[width=\linewidth]{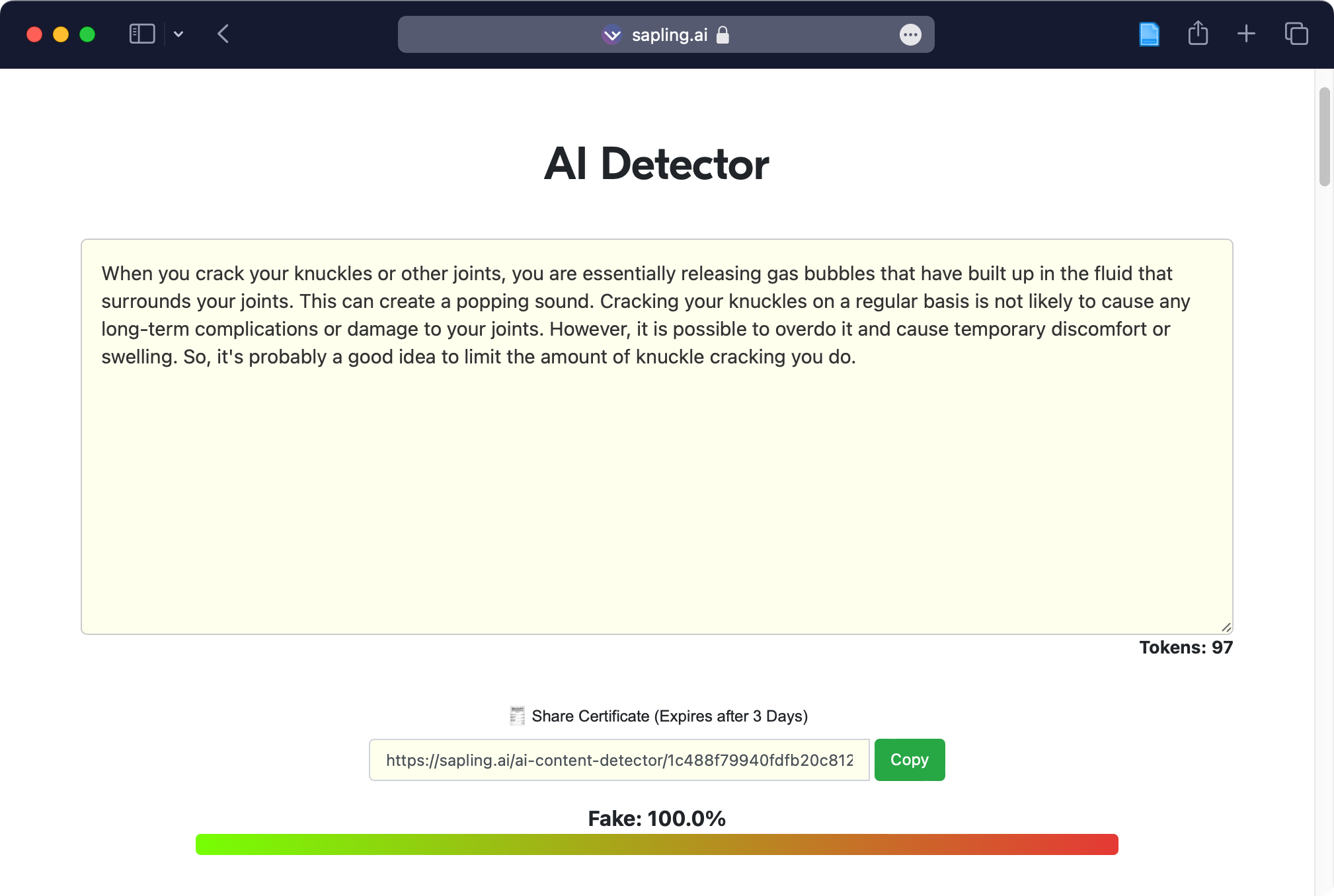}
        \caption{Original text.}
    \end{subfigure}
    \hfill
    \begin{subfigure}{.49\textwidth}
        \centering
        \includegraphics[width=\linewidth]{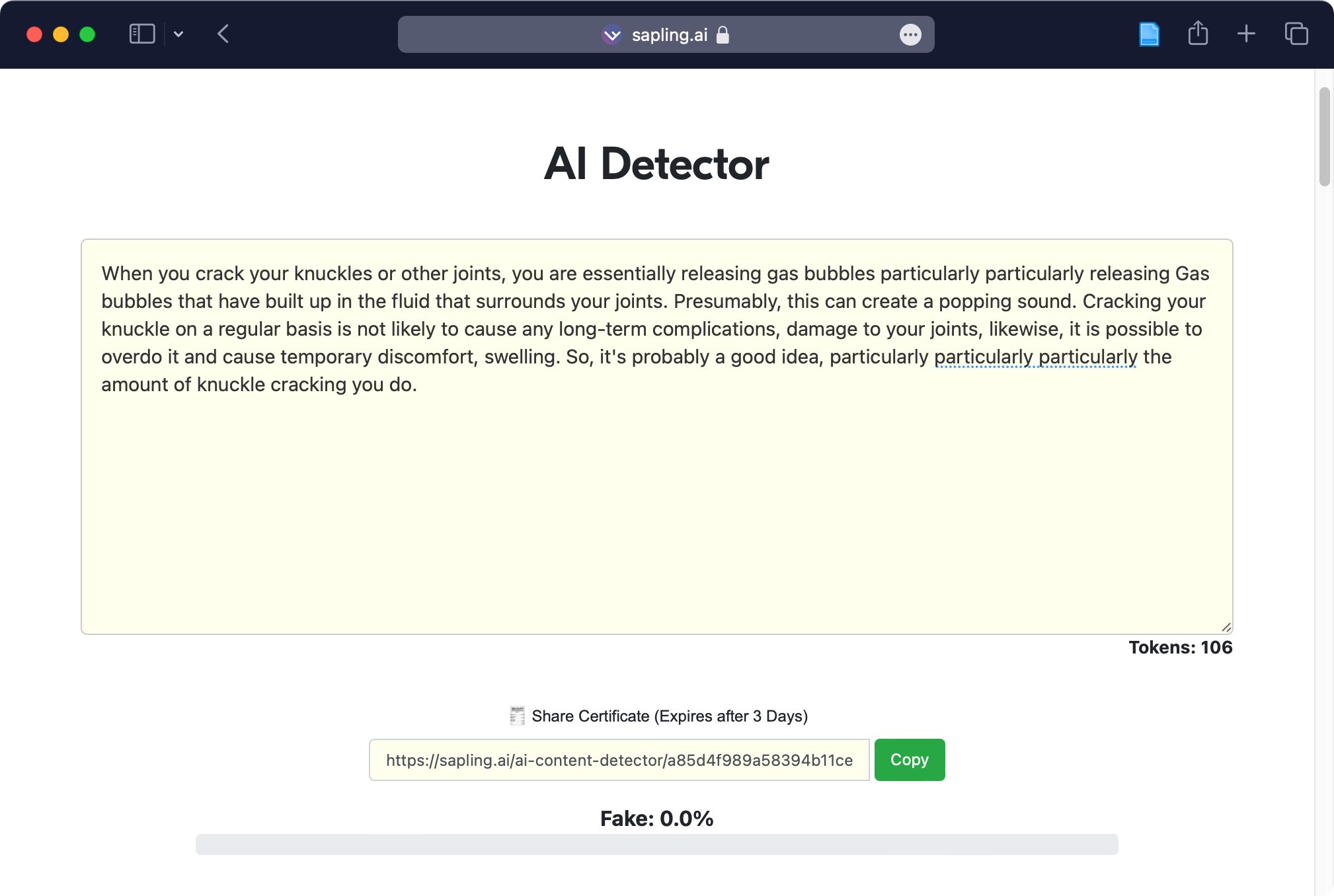}
        \caption{Paraphrased text.}
    \end{subfigure}
    \caption{An attack example of HC3 dataset against Sapling detector.}
    \label{fig:sapling_attack_example}
\end{figure*}

\begin{figure*}[!tbp]
    \centering
    \begin{subfigure}{.49\textwidth}
        \centering
        \includegraphics[width=\linewidth]{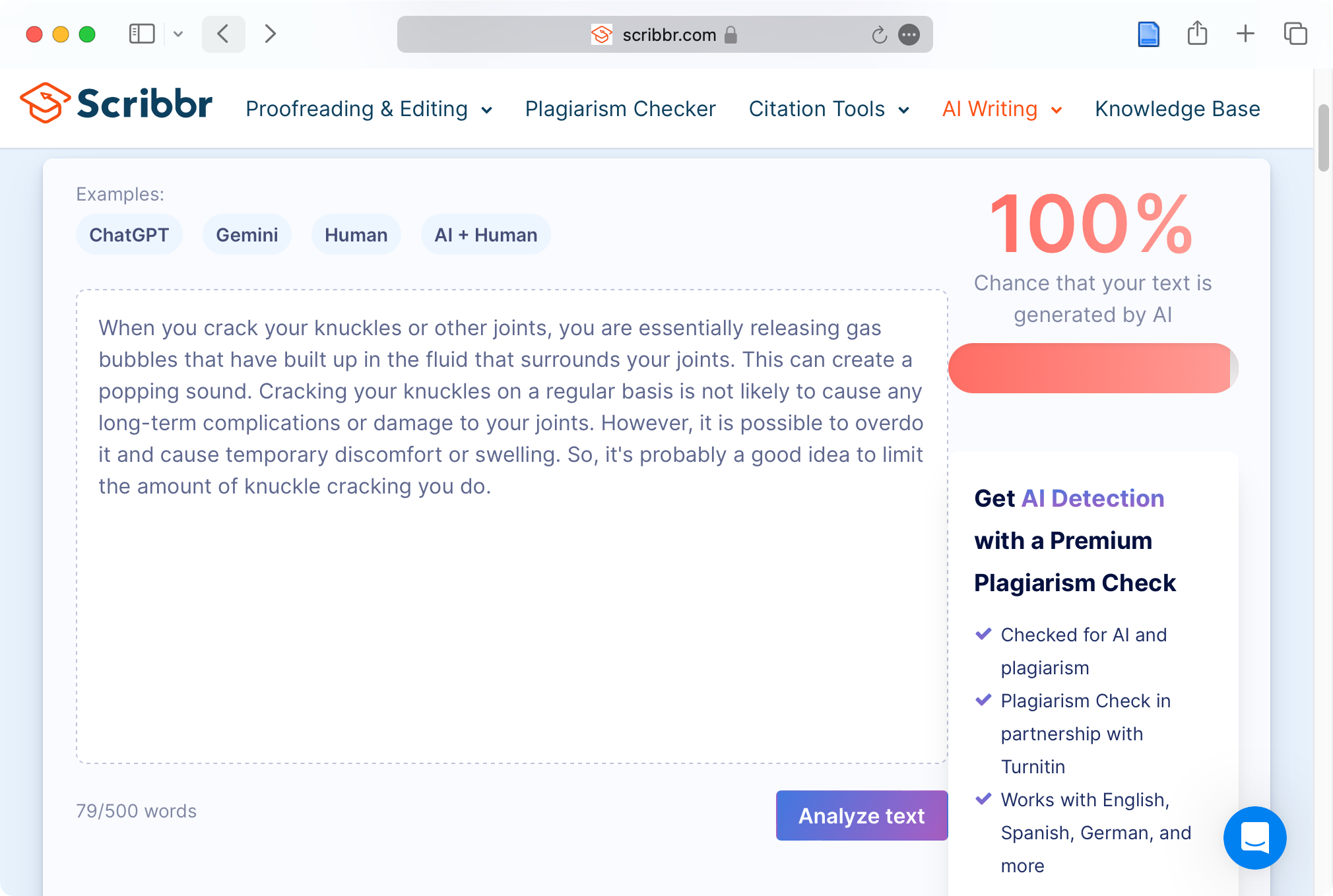}
        \caption{Original text.}
        \label{}
    \end{subfigure}
    \hfill
    \begin{subfigure}{.49\textwidth}
        \centering
        \includegraphics[width=\linewidth]{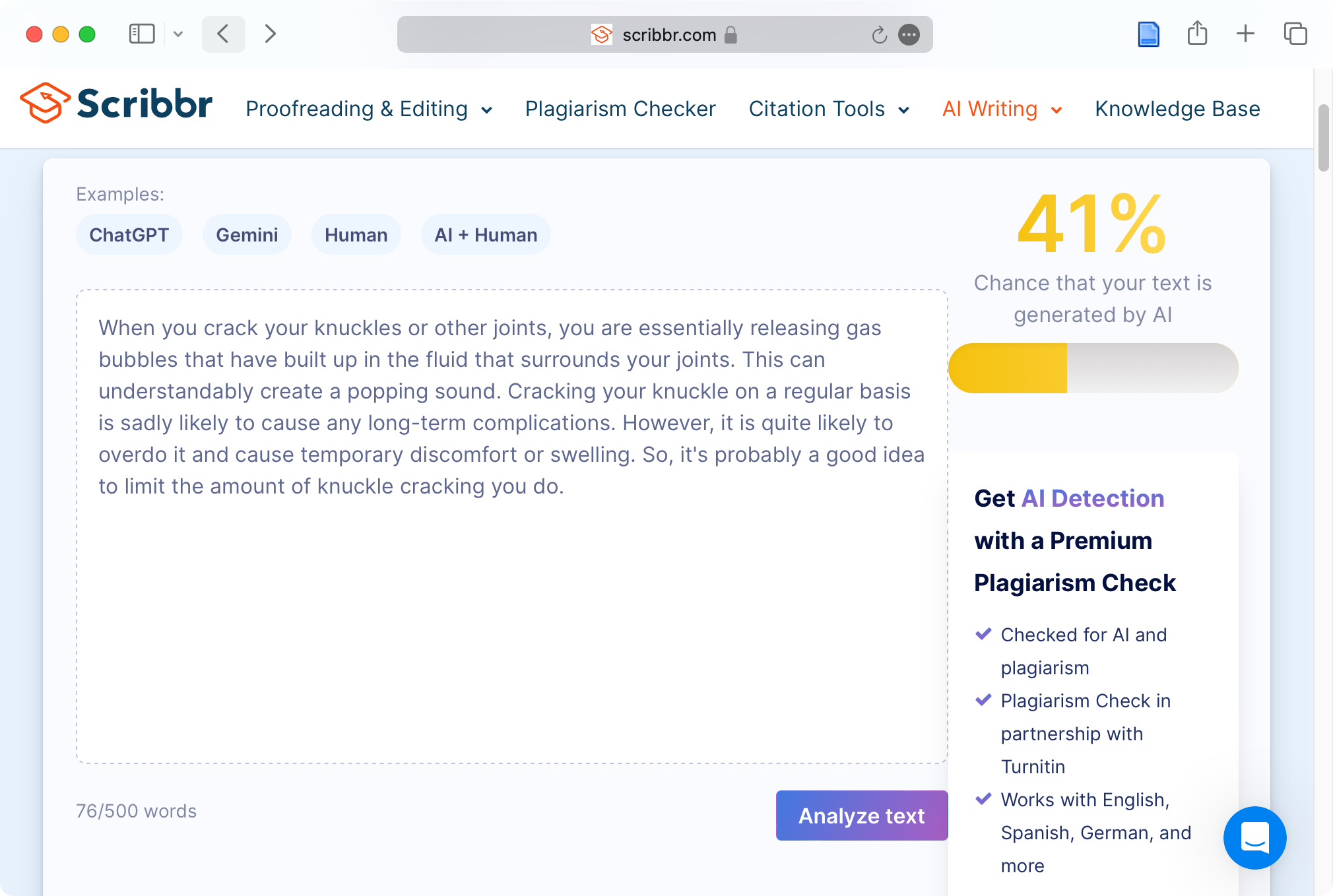}
        \caption{Phraphrased text.}
        \label{}
    \end{subfigure}
    \caption{An attack example of HC3 dataset against Scribbr detector.}
    \label{fig:scribbr_attack_example}
\end{figure*}

\begin{figure}[!tbp]
    \centering
    \includegraphics[width=\columnwidth]{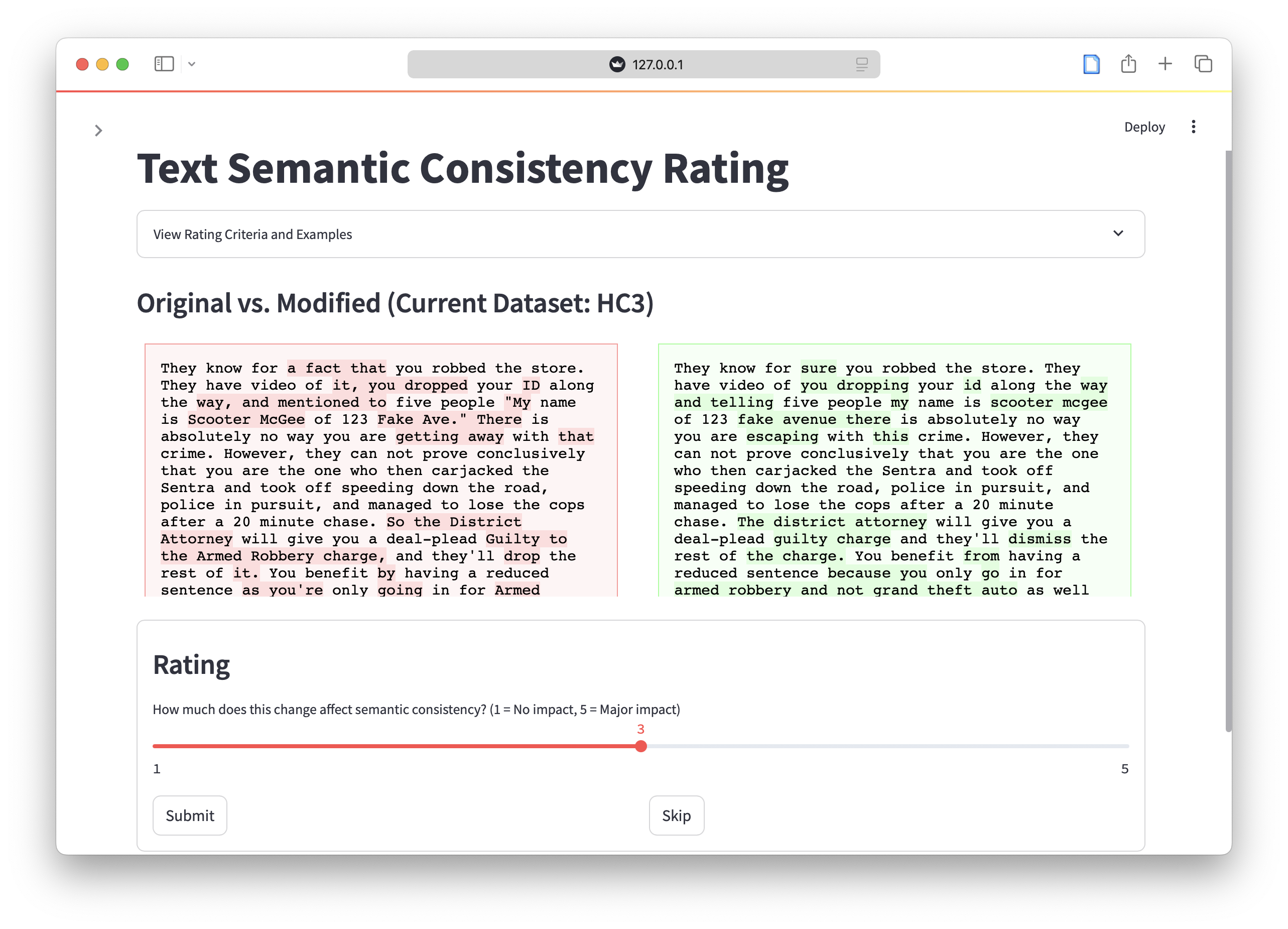}
    \caption{Webpage of text semantic consistency rating platform.}
    \label{fig:streamlit}
\end{figure}

\clearpage
\section{Artifact Appendix}



\subsection{Abstract}
We proposed \method, the first gradient-based evader for attacking AI-generated text (AIGT) detectors. \method overcomes the undifferentiable computation problem, caused by the discrete nature of text, by constructing weighted embeddings for the detector input. It then updates the evader model parameters using feedback from victim detectors, achieving high evasion rates with minimal text modifications.

This artifact is used to evaluate \method against a range of detectors. It includes code for reproducing \method as well as three baseline evaders. We provide a Python library that makes it easy to reproduce existing detectors and evaders. The artifact also contains trained evader models and scripts for replicating our experiments on two real-world AIGT detectors, namely Sapling and Scribbr. To help mitigate the risks posed by AIGT evaders, we showcase that our proposed active paraphrase defense can effectively reduce evasion rates.

\subsection{Description \& Requirements}
We package the artifact into two separate files: \texttt{GradEscape.zip} and \texttt{Usenix-AE.zip}. \texttt{GradEscape.zip} contains the source code; while \texttt{Usenix-AE.zip} contains evaluation datasets and trained models.
This section describes the minimal hardware and software requirements needed to run the artifact.


\subsubsection{Security, privacy, and ethical concerns}
Our attacks target exclusively at the target AIGT detector.
Therefore, there is no security risk for evaluators.
All benchmark datasets are sourced from open corpora, so there are no privacy concerns.
Regarding real-world experiments, Sapling and Scribbr have updated their services with stronger models. Therefore, our artifact poses no significant risk to these two online platforms.

\subsubsection{How to access}
Our artifact is available through Zenodo. The artifact can be accessed at \url{https://doi.org/10.5281/zenodo.15586856}.

\subsubsection{Hardware dependencies}
Two Nvidia RTX A6000 GPUs are the minimum GPU requirement to run the artifact.
We recommend at least a 20-core CPU, 32 GB RAM, and 256 GB of free disk space.

\subsubsection{Software dependencies}
\begin{itemize}
    \item \textbf{OS:} Ubuntu 20.04+. A macOS machine is needed to open Scribbr webarchive file. If you encounter a security warning when opening the webarchive file, please go to Settings, Privacy \& Security and click open anyway.
    \item \textbf{Package Manager:} Conda.
    \item \textbf{API Key:} A Sapling API key is required to run Sapling experiments. Conducting these experiments entails money costs.
\end{itemize}

\subsubsection{Benchmarks}
Our evaluation requires four datasets: GROVER, HC3, GPA, and GPTWiki. GROVER and GPA are provided in \texttt{Usenix-AE.zip}. Our code will automatically download and manage HC3 and GPTWiki. Some pre-trained victim detectors are also included in \texttt{Usenix-AE.zip}.

\subsection{Set-up}
The setup utilizes Conda for environment management.

\subsubsection{Installation}
\mypara{Install Main Environment}
Download \texttt{GradEscape.zip} and \texttt{Usenix-AE.zip}.
Unzip and place them in the same directory.
Then, use the following commands to create an environment:
\begin{verbatim}
conda create -n ge python=3.10
conda activate ge
cd GradEscape
./install.sh
cp src/AIGT/.config.yaml src/AIGT/config.yaml
\end{verbatim}

\mypara{Generate Word Similarity Matrix}
Perturbation-based evaders rely on a word similarity matrix to select synonyms.
\begin{Verbatim}[fontsize=\small,breaklines=true]
cd Usenix-AE
git clone https://github.com/nmrksic/counter-fitting.git
cd counter-fitting/word_vectors/
unzip counter-fitted-vectors.txt.zip
python ../../../GradEscape/tools/comp_cos_sim_mat.py counter-fitted-vectors.txt
\end{Verbatim}
Then edit \texttt{config.yaml} to set the correct data\_dir and counter\_fitting\_path.

\mypara{Create vLLM Environment}
We need vLLM for fast paraphrasing. Since vLLM has complex dependencies, we create a new environment specifically for vLLM.

\begin{verbatim}
conda create -n vllm python=3.10
conda activate vllm
cd GradEscape
pip install -r paraphrase_requirements.txt
\end{verbatim}
Our artifact mainly runs on \texttt{ge}; \texttt{vllm} is only used for paraphrase defense.


\subsubsection{Basic Test}
We provide a simple test that trains a detector using \texttt{AIGT}.
\begin{verbatim}
python basic_test.py
\end{verbatim}
If you see ``Test finished successfully!'', it means the installation was successful.

\subsection{Evaluation workflow}

\subsubsection{Major Claims}

\begin{compactdesc}



    \item[(C1):] \method achieves higher evasion rates than state-of-the-art evaders under the same text quality requirement. This is proven by the experiment (E1) described in Section 6.2 whose results are illustrated in Figure 3 and Figure 4.

    \item[(C2):] \method is effective against real-world commercial AIGT detectors. This is proven by experiment (E2) described in Section 6.6 whose results are reported in Table 4 and illustrated in Figure 20 and Figure 21.

    \item[(C3):] Our proposed potential defense can reduce evasion rates below 0.2 against multiple evaders. This is proven by experiment (E3) described in Section 7 whose results are illustrated in Figure 11.

\end{compactdesc}

\subsubsection{Experiments}

\begin{compactdesc}






    \item[(E1):] \textit{[Verify Evasion Effectiveness] [5 human-minutes + 1 compute-hour]:}

    \begin{asparadesc}
        \item[How to:] Navigate to \texttt{examples} directory; activate \texttt{ge} environment; execute evader training script in \texttt{scripts}. The evasion rate and text quality metrics will be printed on the shell.

        \item[Preparation:] None.

        \item[Execution:]
        Navigate to \texttt{examples} directory and activate \texttt{ge}:
        \begin{verbatim}
cd examples
conda activate ge
        \end{verbatim}
        Then execute the evader training script:
        \begin{verbatim}
./scripts/train_evader_roberta_grover.sh
        \end{verbatim}

        \item[Results:] After the program finishes running, the evasion rate and text quality metrics (perplexity, cosine similarity, GRUEN, and ROUGE) will be printed to the terminal. Evaluators can compare these results with the first row of Figure 3. For ease of reference, we provide details of text quality metrics in~\autoref{tab:metrics}.
    \end{asparadesc}

\begin{table}[htbp]
\centering
\caption{Text quality metrics summary. Expected refers to typical value ranges on GROVER; Print \% indicates whether the metric is displayed as a percentage.}
\label{tab:metrics}
\setlength{\tabcolsep}{3pt}
\resizebox{\columnwidth}{!}{
\begin{tabular}{lcccc}
\toprule
\textbf{Metric} & \textbf{Range} & \textbf{Expected} & \textbf{Larger Better} & \textbf{Print \%} \\
\midrule
Perplexity & $[1, +\infty)$ & $[10, 40]$ & No & No \\
Cos-sim & $[-1, 1]$ & $[0.95, 1.00]$ & Yes & No \\
GRUEN & $[0, 1]$ & $[0.6, 0.8]$ & Yes & No \\
ROUGE & $[0, 1]$ & $[0.8, 1.0]$ & Yes & Yes \\
\bottomrule
\end{tabular}
}
\end{table}

    \item[(E2):] \textit{[Real-world Case Studies] [5 human-minutes + 30 compute-minutes]:}

    \begin{asparadesc}
        \item[How to:] Set your Sapling API key environment parameter; run the two real-world case study Jupyter Notebooks. Evaluators may open \texttt{Scribbr.webarchive} to verify Scribbr results.

        \item[Preparation:] A Sapling API key and a macOS machine for Scribbr verification.

        \item[Execution:]
        Set your Sapling API key:
        \begin{verbatim}
export SAPLING_API_KEY=<your_api_key>
        \end{verbatim}
        Run \texttt{real\_world\_demo\_sapling.ipynb} and \texttt{real\_world\_demo\_scribbr.ipynb}.
        The execution environment is \texttt{ge}.

        \item[Results:] The Sapling results will be printed in its Jupyter Notebook. Evaluators can compare the printed results with Figure 20 and Table 4. Verifying Scribbr results requires copying the output into the website rendered by \texttt{Scribbr.webarchive}.
        The Scribbr results should be the same as Figure 21.
    \end{asparadesc}

    \item[(E3):] \textit{[Paraphrase Defense Experiment] [5 human-minutes + 2 compute-hours]:}

    \begin{asparadesc}
        \item[How to:] Navigate to \texttt{examples} directory. First, use \texttt{vllm} environment to run \texttt{paraphrase\_defense\_grover.sh}. Then, activate \texttt{ge} and run \texttt{eval\_paraphrase\_defense\_grover.sh}. It will generate a figure in the current directory named \texttt{paraphrase\_defense\_grover.pdf}.

        \item[Preparation:] None.

        \item[Execution:] Navigate to \texttt{examples}:
        \begin{verbatim}
cd examples
        \end{verbatim}
        Run paraphrase:
        \begin{verbatim}
conda activate vllm
./scripts/paraphrase_defense_grover.sh
        \end{verbatim}
        Train a new detector and evaluate the defense:
        \begin{verbatim}
conda activate ge
./scripts/eval_paraphrase_defense_grover.sh
        \end{verbatim}

        \item[Results:] The program will generate a figure named \texttt{paraphrase\_defense\_grover.pdf} in \texttt{examples/}. This figure illustrates evasion rates before and after applying our defense. Evaluators can compare the generated figure with Figure 11 in our paper.
    \end{asparadesc}

\end{compactdesc}


\subsection{Notes on Reusability}
\label{sec:reuse}


We implement \texttt{AIGT} in a reusable way. All configurations, including datasets, detectors, and evaders, are extracted in \texttt{arguments.py}. Followers can replicate existing AIGT detectors and evaders with little effort. The user interface of detectors and evaders is designed in a unified and HuggingFace-like way.
Followers can also write their own detectors and evaders in \texttt{detectors/} and \texttt{evaders/}, respectively.


\subsection{Version}
Based on the LaTeX template for Artifact Evaluation V20231005. Submission,
reviewing and badging methodology followed for the evaluation of this artifact
can be found at \url{https://secartifacts.github.io/usenixsec2025/}.

\end{document}